\documentclass[12pt]{article}
\usepackage{UF_FRED_paper_style}
\usepackage{times}
\usepackage{amsmath, amssymb}
\usepackage{geometry}
\usepackage{cancel}
\usepackage{marginnote}
\usepackage{amsthm}
\usepackage{cleveref}
\usepackage{natbib}
\usepackage{appendix}
\usepackage{booktabs}

\hypersetup{
    colorlinks = true,
    citecolor = blue
}
\usepackage{lipsum}  
\usepackage{setspace}
\newtheorem{theorem}{Theorem}
\newtheorem{definition}{Definition}

\usepackage{bm}
\usepackage{booktabs}
\usepackage{subcaption}
\usepackage{graphicx}
\title{\textbf{The Bayesian optimal two-stage design for clinical phase II trials based on Bayes factors}}

\author{Riko Kelter \thanks{Correspondence concerning this article should be addressed to ...
    Draft version 1.0, 28/11/2025. This paper has not been peer reviewed. Please do not copy or cite without author's permission. Data and R code to reproduce our results are openly available at \url{https://osf.io/3dbtv/overview?view_only=9f9d87a6748c446ead76de99c1e830b7}. We declare no conflict of interest.}\\
	Institute of Medical Statistics and Computational Biology\\
	Faculty of Medicine, University of Cologne\\
    \and Samuel Pawel \\
    Epidemiology, Biostatistics and Prevention Institute\\
    Center for Reproducible Science \\
    University of Zurich}

\date{\today}

\begin{document}

{\setstretch{.8}
\maketitle
\begin{abstract}
Sequential trial design is an important statistical approach to increase the efficiency of clinical trials. Bayesian sequential trial design relies primarily on conducting a Monte Carlo simulation under the hypotheses of interest and investigating the resulting design characteristics via Monte Carlo estimates. This approach has several drawbacks, namely that replicating the calibration of a Bayesian design requires repeating a possibly complex Monte Carlo simulation. Furthermore, Monte Carlo standard errors are required to judge the reliability of the simulation. All of this is due to a lack of closed-form or numerical approaches to calibrate a Bayesian design which uses Bayes factors. In this paper, we propose the Bayesian optimal two-stage design for clinical phase II trials based on Bayes factors. The optimal two-stage Bayes factor design is a sequential clinical trial design that is built on the idea of trinomial tree branching, a method we propose to correct the resulting design characteristics for introducing a single interim analysis. We build upon this idea to invent a calibration algorithm which yields the optimal Bayesian design that minimizes the expected sample size under the null hypothesis. Examples show that our design recovers Simon's two-stage optimal design as a special case, improves upon non-sequential Bayesian design based on Bayes factors, and can be calibrated quickly, as it makes use only of standard numerical techniques instead of time-consuming Monte Carlo simulations. Furthermore, the design allows to ensure a minimum probability on compelling evidence in favour of the null hypothesis, which is not possible with other designs. As the idea of trinomial tree branching is neither dependent on the endpoint, nor on the use of Bayes factors, the design can therefore be generalized to other settings, too.


\noindent

\end{abstract}
}

\section{Introduction}\label{sec:intro}

Two-stage designs in clinical trials are primarily used to assess the efficacy of a treatment, especially in early-phase clinical trials where a smaller sample size is typically used \citep{WassmerBrannath2016,Kieser2020}. In these trials, a small group of patients is enrolled and their outcome is evaluated in a first step. After the first stage, an interim analysis is performed which determines whether the trial proceeds to the second stage or is stopped. Only if the interim analysis suggests sufficient evidence of efficacy (of a novel drug or treament), a second stage with additional patients is initiated. There are various benefits of two-stage clinical trial designs, in particular, for single-arm phase IIa trials where the goal is to demonstrate a proof of concept for a novel drug or treament. Among these benefits are:
\begin{itemize}
    \item[$\blacktriangleright$]{\textit{Ethical considerations}: If the treatment shows no promising efficacy in the first stage, the trial can be stopped for futility. This avoids further exposure of patients to potentially ineffective treatment \citep{WassmerBrannath2016}.}
    \item[$\blacktriangleright$]{\textit{Efficiency considerations}: The second stage is only initiated if there is enough statistical evidence of efficacy, which saves resources and time. \citep{Pallmann2018}.}
\end{itemize}

Bayesian approaches to two-stage designs (and, in general, sequential clinical trial designs) are a fast expanding field of research \citep{Stallard2020,Fayers2005,Ferguson2021,Ferreira2021,Zhou2017,Zhou2021,KelterSchnurr2024}. While Bayesian approaches have certain advantages such as the ability to incorporate historical information via the prior distribution, Bayesian two-stage or sequential designs often require a complicated calibration via Monte Carlo simulation studies \citep{Chevret2012}. Calibration refers to guaranteeing a prespecified type-I-error rate and power which must be interpretable from a frequentist point of view \citep{Grieve2022}. This fact is due to health authorities' requirements \citep{FDAAdaptiveBayesianDesignsForClinicalTrials2019,FDAAdaptiveDesignsForClinicalTrials} and complicates the design of Bayesian two-stage and sequential designs. Although Monte Carlo studies allow to simulate trial data under the null and alternative hypothesis $H_0$ and $H_1$ and study the resulting type-I-error and power -- depending on the test criteria used -- the reproducibility of Monte Carlo studies is inherently troubled by various aspects, for details see \cite{Morris2019}, \cite{Kelter2023}, \cite{Seibold2021}, \cite{Boulesteix2020} and \cite{Boulesteix2020a}. Still, calibration of Bayesian two-stage and sequential designs by means of a Monte Carlo simulation is the dominant approach in practice. \citep{Grieve2022,Berry2011,Giovagnoli2021}.\footnote{Regarding the need to carry out possibly complex simulations to calibrate a Bayesian design, \cite[p.~1010]{Chevret2012} notes that a particular \textit{``challenge in using Bayesian adaptive trials is the lack of user-friendly software for study design and analysis. The practical application of these methods has only become feasible since the early 1990s because of computational advances in Markov chain Monte Carlo methods (...).''}} Based on the literature, there is clear lack of non-simulation based approaches to calibrate Bayesian two-stage or sequential trials. This would improve both the reproducibility of the statistical methodology as well as the ease of application of such trial designs.

Furthermore, in such trials, the most popular test criteria is the posterior probability of the hypotheses under consideration \citep{Zhou2023}. Then, the posterior probabilities $P(H_0 \mid y)$ and $P(H_1 \mid y)$ given the data are used to test the null hypothesis $H_0$ versus the alternative $H_1$. However, other options exist: Bayes factors have gained increasing attention recently in the context of two-stage assessment of hypotheses \citep{Schoenbrodt2017,Stefan2022,VandeSchoot2021} as well as in preclinical research \citep{Pourmohammad2023,Kelter2023a}. The Bayes factor $\mathrm{BF}_{01}(y):=\frac{f(y \mid H_0)}{f(y \mid H_1)}$ conceptually is the predictive updating factor from prior to posterior odds:
\begin{align}\label{eq:bayesFactor}
    \underbrace{\frac{P(H_0 \mid y)}{P(H_1 \mid y)}}_{\text{Posterior odds}} = 
    \underbrace{\frac{f(y \mid H_0)}{f(y \mid H_1)}}_{\text{Bayes factor}} \cdot \underbrace{\frac{P(H_0)}{P(H_1)}}_{\text{Prior odds}}
\end{align}
Bayes factors have become increasingly popular in psychology and other areas \citep{VandeSchoot2021}, most probably because they have certain advantages over posterior probabilities. 

First, while posterior probabilities are strongly influenced by the associated prior probabilities $P(H_0)$ and $P(H_1)$ of the hypotheses of interest, Bayes factors are independent of the prior odds of the hypotheses under consideration. Bayes factors are solely influenced by the prior distribution on the parameter itself. As a consequence, even if historical data are used which favour one hypothesis a priori before the trial is conducted, the Bayes factor is not influenced by that prior probability. In contrast, the posterior probability is strongly drawn towards the a priori favoured hypothesis in such a case.

Second, Bayes factors are easy to interpret. For example, a Bayes factor of size ten implies that data are ten times more likely under the null hypothesis $H_0$ than under the alternative $H_1$. This allows for simple communication with clinicians. One could argue that the same holds for posterior probabilities, as probabilities are simple to communicate. However, the posterior probability can be influenced strongly by the prior probabilities. In that case, although the posterior probability scale is easy to communicate, it is often difficult to gauge the influence of the prior probability on the resulting posterior probability.

Third, Bayes factors can be used as a frequentist test statistic and the long-term properties of Bayes factors provide a Bayes-frequentist compromise \citep{Grieve2022,KelterSchnurr2024,VandeSchoot2021,Stallard2020}. We turn to this perspective in the following \Cref{sec:twoStage} in \Cref{eq:bayesianDesign}, our main requirement to calibrate a Bayesian two-stage design. Still, we already stress here that our proposed design can be extended to other testing approaches including posterior probabilities.

Fourth, it is well known that sample size planning and the resulting power under $H_1$ and type-I-error rate under $H_0$ can (and often does) vary for different approaches to testing, such as posterior probabilities and Bayes factors \citep{Kelter2020,Makowski2019,Linde2020ROPE,Kelter2021BMCHodgesLehmann}. Thus, it might be possible that designs which make use of the Bayes factor can improve upon certain operating characteristics of an analogue design which uses posterior probabilities.

Now, an important distinction in Bayesian trial design is between design and analysis priors. While the design prior is selected primarily to incorporate possibly available information and a priori knowledge about the phenomenon under study, the analysis prior is the prior used to calculate the Bayes factor once the data are collected \citep{Berry2011,OHagan2019,Schoenbrodt2017,Stefan2022}. Thus, an analysis prior often needs to be much more objective than a design prior. During the design of a trial, the key requirement is to obtain a calibrated design, which we will explain in further detail in \Cref{sec:twoStage}.

\subsection{Outline}
Although Bayes factors have some appealing properties as outlined above, few clinical trial designs exist which make use of them \citep{Zhou2017,Giovagnoli2021}. In this paper, we develop a Bayesian two-stage clinical trial design for phase IIa trials with binary endpoints. In these trials, we focus on the hypotheses
\begin{align}
    H_0:p\leq p_0 \text{ versus } H_1:p>p_0
\end{align}
for some $p_0\in (0,1)$, so we focus on a proof of concept trial for a novel drug or treatment. Our trial design makes use of Bayes factors to test the above hypotheses, but can conceptually be generalized to any testing approach -- including posterior probabilities and others, see \cite{Kelter2020BayesianPosteriorIndices}. Therefore, we introduce a general graphical structure of our Bayesian two-stage design based on a trinomial tree and provide theoretical results which allows for almost instant computation of the Bayesian power and type-I-error rate of our design. Thus, our approach renders complicated Monte Carlo simulations to calibrate the design entirely obsolete, which is even more appealing as our design can be generalized to other testing approaches, as mentioned above.

The structure of our paper is as follows: In the next \Cref{sec:twoStage} we introduce details about our design and the operating characteristics we focus on when speaking about calibration. \Cref{sec:trinomial} outlines the key problems which must be addressed when allowing for a single interim analysis in the design -- thereby modifying it into a two-stage design. \Cref{sec:solutions} provides solutions to these main problems and \Cref{sec:calibration} then introduces a calibration algorithm based on these solutions. \Cref{sec:examples} provides illustrating examples and shows how the design works in practical settings and \Cref{sec:discussion} closes the paper with a discussion for future work and a conclusion.

\section{The Bayesian two-stage design for phase II clinical trials with binary endpoints}\label{sec:twoStage}

The idea of the two-stage design is as follows: We search for two natural numbers $n_1$ and $n_2$ with $n_1<n_2$ for which the two requirements
\begin{align}\label{eq:bayesianDesign}
    P(\mathrm{BF}_{01}^{n_2}(y) < k \mid H_0) \leq \alpha \hspace{1cm} \text{and} \hspace{1cm} P(\mathrm{BF}_{01}^{n_2}(y) < k \mid H_1) \geq 1-\beta_{n_2}
\end{align}
are met for some $\alpha$ and $\beta_{n_2}$ in $(0,1)$. The first denotes the Bayesian type-I-error rate, while the second resembles a Bayesian analogue of frequentist power.
For example, for $k=1/10$ -- the usual threshold for strong evidence according to the scale of \cite{Jeffreys1939} -- when $P(\mathrm{BF}_{01}^{n_2}(y) < k \mid H_0)\leq \alpha$ holds, the probability to obtain a Bayes factor indicating strong evidence against the null hypothesis (that is, in favour of the alternative $H_1$) is bounded by $\alpha$ under $H_0$.\footnote{This holds because $P(\mathrm{BF}_{01}^{n_2}(y) < k \mid H_0)=P(1/\mathrm{BF}_{01}^{n_2}(y) > 1/k \mid H_0)=P(\mathrm{BF}_{10}^{n_2}(y) >10 \mid H_0) \leq \alpha$ and $\mathrm{BF}_{10}^{n_2}(y) > 10$ amounts to the Bayes factor in favour of $H_1$ indicating (at least) strong evidence for $H_1$. Thus, $P(\mathrm{BF}_{01}^{n_2}(y) < k \mid H_0)$ resembles an intuitive Bayesian type-I-error rate.} Thus, for $\alpha:=0.05$ the Bayesian type-I-error rate is controlled at $5\%$ then. Likewise, if $P(\mathrm{BF}_{01}^{n_2}(y) < k \mid H_1)>1-\beta_{n_2}$ for, say $\beta_{n_2}:=0.2$ and $k=1/10$, the probability to obtain strong evidence in favour of $H_1$, when $H_1$ is true, is $80\%$.\footnote{This holds because $P(\mathrm{BF}_{01}^{n_2}(y) < k \mid H_1)=P(\mathrm{BF}_{10}^{n_2}(y) > 1/k \mid H_1)=P(\mathrm{BF}_{10}^{n_2}(y) > 10 \mid H_1)$.} A critical goal in Bayesian design of experiments and sample size calculation therefore is to provide a solution to the inequalities in \Cref{eq:bayesianDesign}
for some $\alpha$ and $\beta_{n_2}$ in $(0,1)$.

In \Cref{eq:bayesianDesign}, the superscript $n_2$ in $\mathrm{BF}_{01}^{n_2}(y)$ indicates that the Bayes factor is based on a sample of size $n_2$. The sample size $n_2$ is the sample size at the end of the trial after $n_2$ patients have been enrolled and the outcomes have been observed. 

\begin{figure}[h!]
    \centering
    \includegraphics[width=1\linewidth]{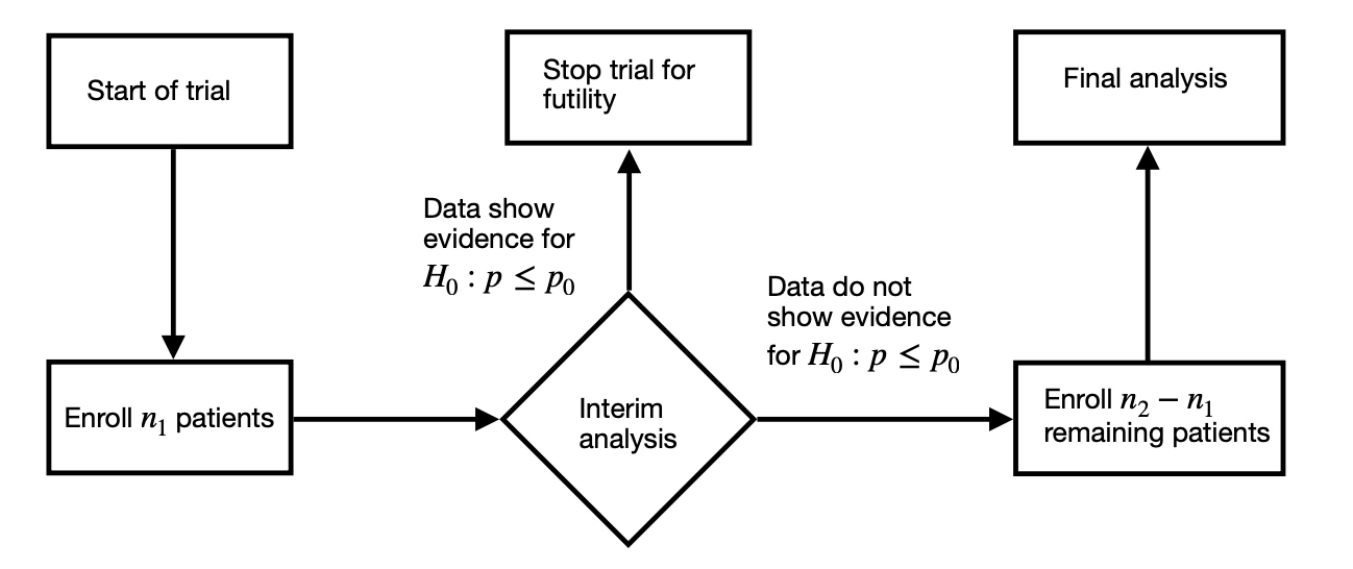}
    \caption{Overview of the Bayesian two-stage design for phase IIa clinical trials}
    \label{fig:1}
\end{figure}

The structure of the Bayesian single-arm two-stage design with binary endpoints is shown in \Cref{fig:1}. The two-stage design makes a single interim analysis after the outcomes of $n_1$ patients have been enrolled. The principal goal is to stop the trial for futility at this interim analysis, if there is sufficient evidence for the null hypothesis $H_0:p\leq p_0$ that the drug or novel treatment is ineffective. Here, $p_0$ denotes a success probability, and if the true success probability of the drug is smaller or equal to $p_0$, we deem the drug ineffective (possibly compared to a standard treatment of care with efficacy $p_0$). Thus, the hypotheses of interest are the usual ones considered during a single-arm phase IIa clinical trial -- a proof of concept trial -- where we test
\begin{align}
    H_0:p\leq p_0 \text{ versus } H_1:p>p_0
\end{align}
for some $p_0\in (0,1)$, which corresponds to the efficacy of the standard of care. The primary endpoint is therefore supposed to be binary and measure success or failure of the novel drug or treatment.

As \Cref{fig:1} shows, there is no option to stop the trial for efficacy -- in that case, the trial is continued until reaching the final sample size $n_2$, which is a common choice in phase II trials \citep{Lee2008,Berry2011} -- but it can be terminated early due to futility after observing $n_1$ outcomes. After observing $n_1$ outcomes, the interim analysis is performed to possibly stop the trial early for futility. As a consequence, the observed data $y$ in $\mathrm{BF}_{01}^{n_2}(y)$ consist of the first $y^1:=(y_1,...,y_{n_1})$ observations, and the remaining $y^2=(y_{n_1+1},...,y_{n_2})$ observations, where $y_i=0$ denotes a failure and $y_i=1$ a success for $i=1,...,n_2$.

\subsection{Operating characteristics}
First and foremost, the two-stage design needs to meet the requirements in \Cref{eq:bayesianDesign} regarding Bayesian power and type-I-error rate. We will, see, however, that it is possible to make these error rates fully frequentist under a specific design prior choice, which we will discuss later in detail. Next, to the conditions in \Cref{eq:bayesianDesign}, there are two further operating characteristics which primarily are involved with the interim analysis. In the interim analysis, two criteria or operating characteristics of the trial design can be considered. First, suppose $H_0$ indeed holds. If that's the case, then it seems natural to require a minimum probability
\begin{align}\label{eq:futilityInterim}
    P(\mathrm{BF}_{01}^{n_1}(y^1)>k_f|H_0)>f
\end{align}
for some $f\in (0,1)$, which can be interpreted as guaranteeing a minimum probability $f$ to stop for futility when the drug is not effective and $H_0$ holds. For example, if $f=0.6$, we would stop at the interim analysis with 60\% probability when the drug is ineffective and $H_0:p\leq p_0$ holds. Note that we explicitly use the threshold $k_f$ instead of $k$, because one could use different evidence thresholds for the interim analysis to stop for futility and for the final analysis at the end of the trial, compare \Cref{eq:bayesianDesign}.



We sum up the above and clarify the goal of the Bayesian two-stage design now. 
\begin{itemize}
    \item[$\blacktriangleright$]{We search for two sample sizes $n_1,n_2\in \mathbb{N}$ with $n_1<n_2$, for which \Cref{eq:bayesianDesign} holds for prespecified values of $\alpha\in (0,1)$ and $\beta_{n_2}\in(0,1)$, and for which \Cref{eq:futilityInterim} holds for a prespecified $f\in (0,1)$.}
\end{itemize}

\section{Trinomial tree branching and the problems with power and error rates}\label{sec:trinomial}
In this section, we introduce our approach to calculate the operating characteristics in \Cref{eq:bayesianDesign} for the Bayesian two-stage design described in the previous section. Therefore, we assume a fixed sample size $n_2$ at the final analysis and strive for a correct calculation of the Bayesian power $P(\mathrm{BF}_{01}^{n_2}(y) < k \mid H_1)$ in \Cref{eq:bayesianDesign} first. 

Therefore, we provide the structure of the design graphically in form of a trinomial tree with its different branches. Then, we analyze two core problems which appear when introducing the option to stop the trial for futility after a single interim analysis at $n_1$. These two problems are:
\begin{itemize}
    \item[$\blacktriangleright$]{The (Bayesian) power $P(\mathrm{BF}_{01}^{n_2}(y) < k \mid H_1)$ is too large and must be adjusted due to the introduced interim analysis.}
    \item[$\blacktriangleright$]{The same holds for the (Bayesian) type-I-error rate $P(\mathrm{BF}_{01}^{n_2}(y) < k \mid H_0)$, which is, as it turns out, too conservative and even improves when being adjusted correctly, due to the possibility to stop the trial for futility at the interim analysis.\footnote{Although surprising at first, this fact makes intuitively sense, because allowing to stop for futility at $n_1$ implies that under $H_0$, the Bayes factor cannot be swayed around anymore to show evidence for $H_1$. Thus, after $n_1$ patients the Bayes factor could signal evidence for $H_0$, but after enrolling the remaining $n_2-n_1$ patients show evidence for $H_1$, resulting in a false-positive. If an interim analysis allows to stop at $n_1$ now, the type-I-error rate decreases, as these cases cannot happen anymore. A crucial requirement here is that we do not allow to stop for efficiency, which would influence the type-I-error rate, too.}}
\end{itemize} 
In the following \Cref{sec:solutions}, we then turn to solving both of these issues. This allows to calculate the power and type-I-error rate for the Bayesian two-stage design correctly for a fixed sample size $n_2$. In the subsequent \Cref{sec:calibration}, we then turn to the task of calibrating the Bayesian two-stage design based on our solution. That is, we turn to the issue of providing values $n_1$ and $n_2$ for which \Cref{eq:bayesianDesign} and \Cref{eq:futilityInterim} hold. In this step, we make use of our solution for the correct power and type-I-error rate calculations for a fixed sample size $n_2$ derived in \Cref{sec:solutions}.

\subsection{Graphical structure of the design: The trinomial tree}
\begin{figure}[h!]
    \centering
    \includegraphics[width=1\linewidth]{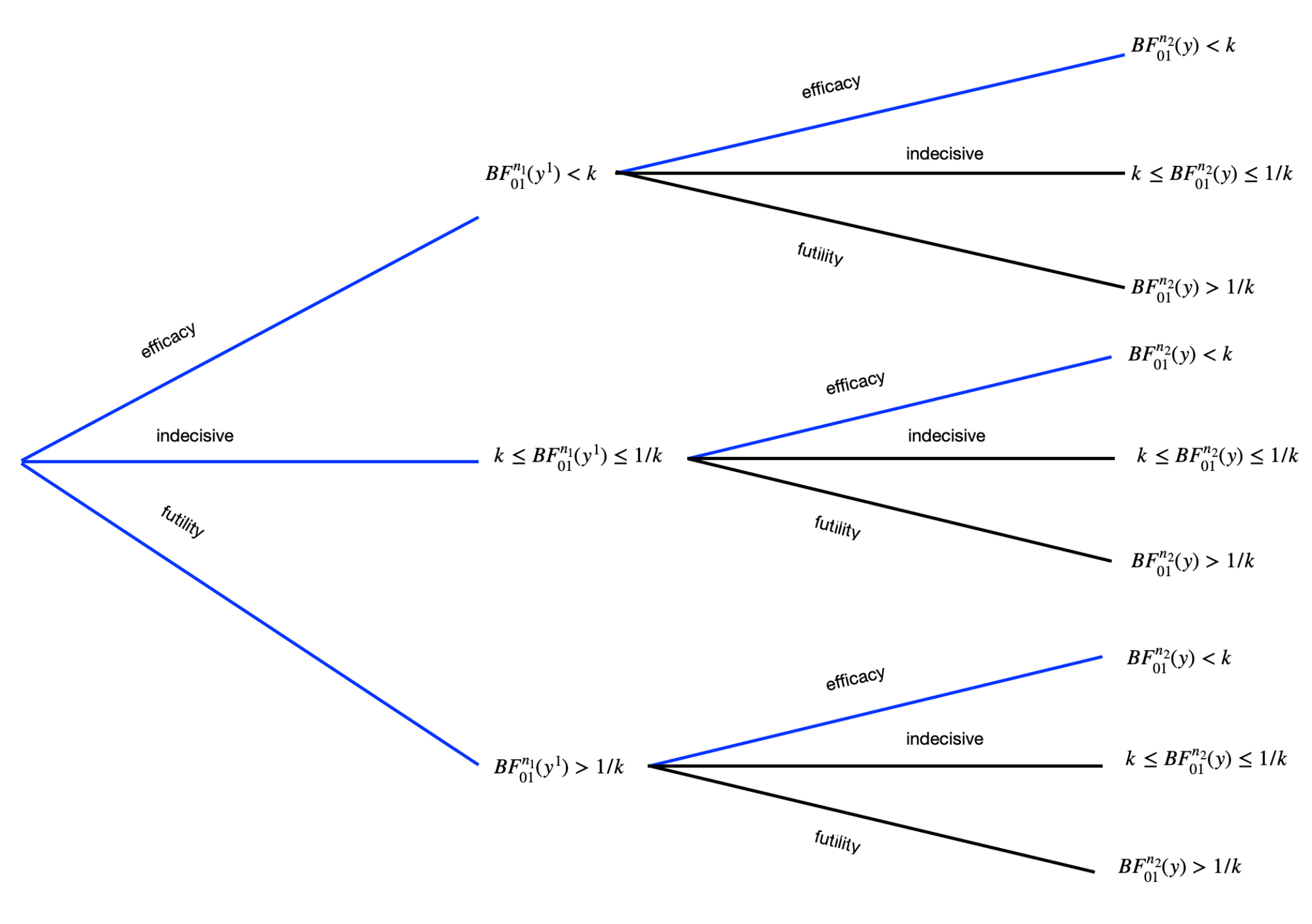}
    \caption{A trinomial tree illustrating the three outcomes the Bayes factor (or a statistical test, in general) can yield during the interim analysis and the final analysis of the Bayesian two-stage design.}
    \label{fig:treePower}
\end{figure}
\Cref{fig:treePower} shows a trinomial tree which accounts for the three possibilities to
\begin{itemize}
    \item{obtain a convincing Bayes factor $\mathrm{BF}_{01}^{n_1}(y^1)<k$, which shows evidence for efficacy of the drug, that is, for $H_1:p>p_0$}
    \item{obtain an indecisive Bayes factor $k\leq \mathrm{BF}_{01}^{n_1}(y^1)\leq 1/k$ which shows neither convincing evidence for efficacy or futility, that is, neither for $H_0:p\leq_0$ or $H_1:p>p_0$}
    \item{obtain an unconvincing Bayes factor $\mathrm{BF}_{01}^{n_1}(y^1)>1/k$ which shows evidence for inefficacy of the drug, that is, for $H_0:p\leq p_0$, and therefore for stopping the trial due to futility.}
\end{itemize}
These three possibilities are possible both at the interim and at the final analysis. The superscripts $n_1$ at the interim Bayes factors in the middle of \Cref{fig:treePower} indicate that we use only the first batch of sample size $n_1$ to compute the latter. The final Bayes factors on the right-hand side of the trinomial tree in \Cref{fig:treePower} make use of all data, that is, of the full sample of sample size $n_2$. All probabilities -- that is, the labels 'efficacy', 'indecisive' and 'convincing' in \Cref{fig:treePower} -- are conditional on $H_i$, so different branches of the trinomial tree have different interpretations when $H_i=H_0$ and when $H_i=H_1$. For example, suppose $H_i=H_1$. Then, the Bayesian power in \Cref{eq:bayesianDesign} is comprised of three branches in total: These are the three blue branches which eventually end in $\mathrm{BF}_{01}^{n_2}(y) < k$ at the right-hand side of \Cref{fig:treePower}. While it seems obvious to calculate power as the sum of the probabilities associated with the three blue branches, there is a caveat: When proceeding like that, the resulting (Bayesian) power is overestimating the true power which results for the two-stage design. We turn to this first problem now.

\subsection{Problem I - Calculating the correct power}
\begin{figure}[h!]
    \centering
    \includegraphics[width=1\linewidth]{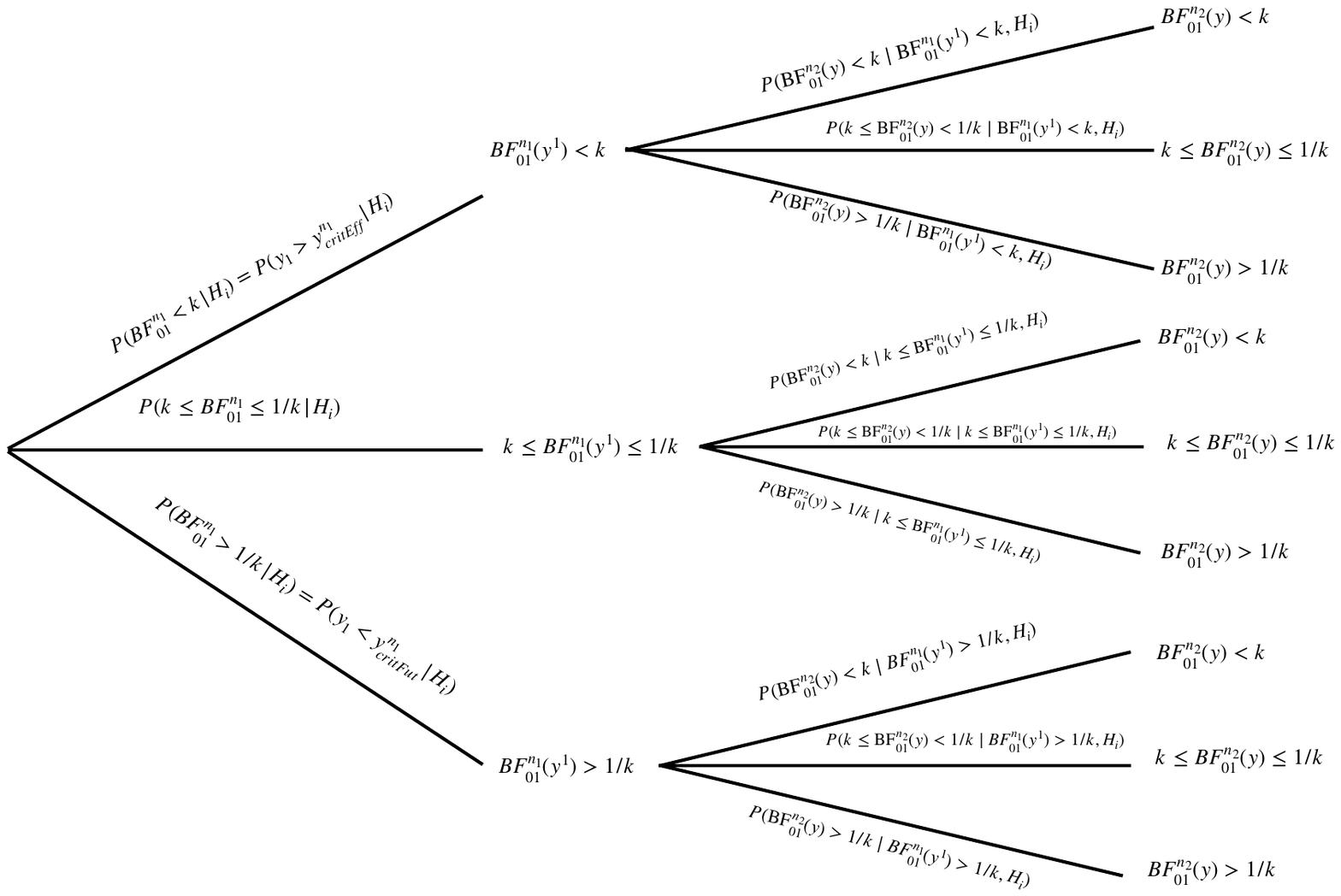}
    \caption{A trinomial tree illustrating the three outcomes the Bayes factor (or a statistical test, in general) can yield during the interim analysis and the final analysis of the Bayesian two-stage design. In addition to \Cref{fig:treePower}, the probabilities of each branch are added.}
    \label{fig:twoStageDesign}
\end{figure}

We first inspect the Bayesian power $P(\mathrm{BF}_{01}^{n_2}(y) < k \mid H_1)> 1-\beta_{n_2}$ based on \Cref{fig:twoStageDesign}, which shows a more detailed graphical structure of the design and is used to explain the first problem in what follows. Based on \Cref{fig:twoStageDesign}, we expand $P(\mathrm{BF}_{01}^{n_2}(y) < k \mid H_1)$ as
\begin{align}\label{eq:conditional}
    P(\mathrm{BF}_{01}^{n_2}(y)& < k \mid H_1)=\underbrace{P(\mathrm{BF}_{01}^{n_2}(y) < k \mid \mathrm{BF}_{01}^{n_1}(y^1)<k,H_1)}_{\text{Efficacy at final analysis given efficacy at interim}}\cdot \underbrace{P(\mathrm{BF}_{01}^{n_1}(y^1)<k\mid H_1)}_{\text{Efficacy at interim analysis}}\nonumber\\
    &+ \underbrace{P(\mathrm{BF}_{01}^{n_2}(y) < k \mid k\leq \mathrm{BF}_{01}^{n_1}(y^1)<1/k,H_1)}_{\text{Efficacy at final analysis given indecisive result at interim}}\cdot \underbrace{P(k\leq \mathrm{BF}_{01}^{n_1}(y^1)<1/k\mid H_1)}_{\text{Indecisive result at interim analysis}}\nonumber\\
    &+ \underbrace{P(\mathrm{BF}_{01}^{n_2}(y) < k \mid \mathrm{BF}_{01}^{n_1}(y^1)>1/k,H_1)}_{\text{Efficacy at final analysis given futility at interim}}\cdot \underbrace{P(\mathrm{BF}_{01}^{n_1}(y^1)>1/k\mid H_1)}_{\text{Futility at interim analysis}}
\end{align}
where the right-hand side equals the probabilities calculated via the path rule in \Cref{fig:twoStageDesign} for the three paths which lead to a $\mathrm{BF}_{01}^{n_2}(y) < k$ after $n_2$ patients have been observed. These are precisely the paths which are shown in blue in \Cref{fig:treePower}, and which end with a Bayes factor indicating efficacy of the novel drug irrespective of what the Bayes factor $\mathrm{BF}_{01}^{n_1}(y^1)$ indicated at the interim analysis.

Now, suppose we find a value $n_2$ for which $P(\mathrm{BF}_{01}^{n_2}(y) < k \mid H_1)> 1-\beta_{n_2}$ for the chosen $\beta_{n_2}$ now. Then this unconditional probability $P(\mathrm{BF}_{01}^{n_2}(y) < k \mid H_1)$ consists of the three probabilities in the right-hand side of 
\Cref{eq:conditional}. However, as we can stop the trial after $n_1$ observed outcomes due to futility at interim, it is possible that the last of these three probabilities, -- the lowest blue branch in \Cref{fig:treePower} -- namely
\begin{align}\label{eq:partialPower}
    \underbrace{P(\mathrm{BF}_{01}^{n_2}(y) < k \mid \mathrm{BF}_{01}^{n_1}(y^1)>1/k,H_1)}_{\text{Efficacy at final analysis given futility at interim}}\cdot \underbrace{P(\mathrm{BF}_{01}^{n_1}(y^1)>1/k\mid H_1)}_{\text{Futility at interim analysis}}
\end{align}
cannot contribute to this power. Conceptually, in this case the Bayes factor provides convincing evidence for $H_0$ at the interim analysis, but if the trial would continue it would sway around and indicate convincing evidence for $H_1$ at the end of the trial, thus contributing to the total power $P(\mathrm{BF}_{01}^{n_2}(y) < k \mid H_1)$. However, as we allow the option to stop the trial for futility, once $\mathrm{BF}_{01}^{n_1}(y^1)>1/k$ holds at the interim analysis the trial is stopped for futility. Thus, the  branch with the probability in \Cref{eq:partialPower} is 'pruned' in a metaphoric sense. Therefore, when calculating unconditional power $P(\mathrm{BF}_{01}^{n_2}(y) < k \mid H_1)$, we must \textit{subtract} the partial power in \Cref{eq:partialPower} from the Bayesian power $P(\mathrm{BF}_{01}^{n_2}(y) < k \mid H_1)$ to obtain the true power of the resulting two-stage design, compare \Cref{eq:conditional}. This adjustment accounts for the possibility that we stop the trial at the interim analysis due to futility. The adjusted Bayesian power therefore reduces to:
\begin{align}\label{eq:adjustedBayesianPower}
    P(\mathrm{BF}_{01}^{n_2}(y)& < k \mid H_1)=\underbrace{P(\mathrm{BF}_{01}^{n_2}(y) < k \mid \mathrm{BF}_{01}^{n_1}(y^1)<k,H_1)}_{\text{Efficacy at final analysis given efficacy at interim}}\cdot \underbrace{P(\mathrm{BF}_{01}^{n_1}(y^1)<k\mid H_1)}_{\text{Efficacy at interim analysis}}\nonumber\\
    &+ \underbrace{P(\mathrm{BF}_{01}^{n_2}(y) < k \mid k\leq \mathrm{BF}_{01}^{n_1}(y^1)<1/k,H_1)}_{\text{Efficacy at final analysis given indecisive result at interim}}\cdot \underbrace{P(k\leq \mathrm{BF}_{01}^{n_1}(y^1)<1/k\mid H_1)}_{\text{Indecisive result at interim analysis}}\nonumber\\
    &\bcancel{+ \underbrace{P(\mathrm{BF}_{01}^{n_2}(y) < k \mid \mathrm{BF}_{01}^{n_1}(y^1)>1/k,H_1)}_{\text{Efficacy at final analysis given futility at interim}}}\cdot \bcancel{\underbrace{P(\mathrm{BF}_{01}^{n_1}(y^1)>1/k\mid H_1)}_{\text{Futility at interim analysis}}}
\end{align}

\subsection{Problem II - Calculating the type-I-error rate}
Likewise, suppose now that $H_0$ holds. In that case, it follows analogue from \Cref{eq:partialPower} with $H_1$ replaced by $H_0$ that stopping for futility implies that no type-I-error can occur once we stop for futility. In that case, the Bayes factor indicates that we stop the trial for futility after $n_1$ observed outcomes, but if the trial would continue the Bayes factor could eventually sway around and indicate evidence for $H_1$, contributing to the type-I-error rate. Thus, when computing the Bayesian type-I-error rate unconditionally as 
$$P(\mathrm{BF}_{01}^{n_2}(y) < k \mid H_0)\leq \alpha$$
compare \Cref{eq:bayesianDesign}, we \textit{overestimate} the true Bayesian type-I-error rate. As a consequence, when allowing to stop the trial for futility at the interim analysis, we must \textit{subtract} the probability in \Cref{eq:partialPower} with $H_1$ replaced by $H_0$ from the Bayesian type-I-error rate $P(\mathrm{BF}_{01}^{n_2}(y) < k \mid H_0)$ to obtain the correct Bayesian false-positive rate. The latter therefore reduces to:
\begin{align}\label{eq:adjustedTypeIErrorRate}
    P(\mathrm{BF}_{01}^{n_2}(y) &< k \mid H_0)=\underbrace{P(\mathrm{BF}_{01}^{n_2}(y) < k \mid \mathrm{BF}_{01}^{n_1}(y^1)<k,H_0)}_{\text{Efficacy at final analysis given efficacy at interim}}\cdot \underbrace{P(\mathrm{BF}_{01}^{n_1}(y^1)<k\mid H_0)}_{\text{Efficacy at interim analysis}}\nonumber\\
    &+ \underbrace{P(\mathrm{BF}_{01}^{n_2}(y) < k \mid k\leq \mathrm{BF}_{01}^{n_1}(y^1)<1/k,H_0)}_{\text{Efficacy at final analysis given indecisive result at interim}}\cdot \underbrace{P(k\leq \mathrm{BF}_{01}^{n_1}(y^1)<1/k\mid H_0)}_{\text{Indecisive result at interim analysis}}\nonumber\\
    &\bcancel{+ \underbrace{P(\mathrm{BF}_{01}^{n_2}(y) < k \mid \mathrm{BF}_{01}^{n_1}(y^1)>1/k,H_0)}_{\text{Efficacy at final analysis given futility at interim}}}\cdot \bcancel{\underbrace{P(\mathrm{BF}_{01}^{n_1}(y^1)>1/k\mid H_0)}_{\text{Futility at interim analysis}}}
\end{align}
Otherwise, we overestimate the latter due to the option to stop the trial for futility after $n_1$ observed outcomes.\footnote{See also footnote 4. If we solely base the type-I-error control on the probability in \Cref{eq:bayesianDesign}, we guarantee that the latter is fully controlled. However, we are too conservative in the sense that we overestimate the true type-I-error rate if we do not adjust it for introducing the option to stop the trial for futility, compare \Cref{eq:adjustedTypeIErrorRate}.}

Summing up, the price paid in statistical terms for introducing an interim analysis in the Bayesian two-stage design is (1) an adjustment which is needed to correct the Bayesian power based on sample size $n_2$ and (2) an adjustment which is needed to correct the Bayesian type-I-error rate based on sample size $n_2$. In the following section, we turn to a solution of both of these issues. This allows to correctly calculate the Bayesian power and type-I-error rate of the Bayesian two-stage design in \Cref{fig:1} for a fixed sample size $n_2$, thereby meeting the criteria in \Cref{eq:bayesianDesign}.

\section{Solving both problems}\label{sec:solutions}
Now, from a computational point of view it is straightforward to obtain the unadjusted Bayesian power $P(\mathrm{BF}_{01}^{n_2}(y)< k \mid H_1)$ or type-I-error rate $P(\mathrm{BF}_{01}^{n_2}(y)< k \mid H_0)$ for a fixed sample size $n_2$. The principal approach to obtain such a power is detailed in \cite{PawelHeld2024}, and details for a single-arm phase IIa trial are provided in \cite{KelterPawel2025}.

Thus, subtracting the probability in \Cref{eq:partialPower} from the unconditional (Bayesian) power requires first to compute that probability. Then, one can calculate the unadjusted Bayesian power and subtract the probability in \Cref{eq:partialPower}, see \Cref{eq:adjustedBayesianPower}. The probability in \Cref{eq:partialPower} can be rewritten as
\begin{align}
    &\underbrace{P(\mathrm{BF}_{01}^{n_2}(y) < k \mid \mathrm{BF}_{01}^{n_1}(y^1)>1/k,H_1)}_{\text{Efficacy at final analysis given futility at interim}}\cdot \underbrace{P(\mathrm{BF}_{01}^{n_1}(y^1)>1/k\mid H_1)}_{\text{Futility at interim analysis}}\\
    &=\underbrace{P(\mathrm{BF}_{01}^{n_2}(y) < k , \mathrm{BF}_{01}^{n_1}(y^1)>1/k \mid H_1)}_{\text{Efficacy at final analysis and futility at interim = futility-erased partial power}}
\end{align}
which we call henceforth \textit{futility-erased partial power}. 
The following result provides a convenient numerical way to calculate $P(\mathrm{BF}_{01}^{n_2}(y) < k , \mathrm{BF}_{01}^{n_1}(y^1)>1/k \mid H_1)$:

\begin{theorem}\label{theorem:1}
Let $n_1$ and $n_2$ be the sample sizes of the interim and final analysis in the Bayesian two-stage design. Let the data $Y_i \mid \theta \sim \mathrm{Bin}(n_i, \theta)$ for $i\in\{1,2\}$ and assume the success probability $\theta$ has a truncated Beta prior, $\theta \sim \mathrm{Beta}(a,b)_{[l,u]}$. Denote by $f(y_s^2,y_s^2|a_d,b_d,l,u)$ the prior-predictive density under the hypothesis $H_i$, for $i=0,1$, where $H_0:p\leq p_0$ and $H_1:p>p_0$ for some $p_0\in (0,1)$. Denote $y_{critEff}^{n_2}$ as the critical value so that for $y_s^2>y_{critEff}^{n_2}$ we have $\mathrm{BF}_{01}^{n_2}(y) < k \mid H_i$ based on the prior-predictive. Denote $y_{critFut}^{n_1}$ as the critical value so that for $y_s^1\leq y_{critFut}^{n_1}$ we have $\mathrm{BF}_{01}^{n_1}(y^1) > 1/k \mid H_i$, where $y_s^1$ denote the number of successes in the first batch $y^1:=(y_1,...,y_{n_1})$ of patients and $y_s^2$ the number of successes in the second batch $y^2:=(y_{n_1+1},...,n_2)$ of patients to be possibly enrolled in the second stage of the trial. Then, the following equality holds: 
    \begin{align}\label{eq:theorem1}
        P(\mathrm{BF}_{01}^{n_2}<k,\mathrm{BF}_{01}^{n_1}>1/k \mid H_i)=\sum_{i=0}^{\lfloor y_{critFut}^{n_1}\rfloor}\sum_{j=\lceil y_{critEff}^{n_2}\rceil -y_i}^{n_2-n_1} f(y_s^1=y_i, y_s^2=y_j|a_d,b_d,l,u)
    \end{align}
\end{theorem}
\begin{proof}
See the Appendix.
\end{proof}

Now, the denominator in \Cref{theorem:1} can be obtained immediately via the Bayes factor sample size planning with root-finding as outlined by \cite{KelterPawel2025}. It is equal to the prior-predictive probability of $y_1<y_{critFut}^{n_1}$ under $H_i$ for the fixed sample size $n_1$.

The numerator is more complicated and is a sum over the joint marginal probability mass function. However, as both data in the first batch of size $n_1$ and in the second batch of size $n_2$ is binomially distributed, the left-hand side probability in \Cref{theorem:1} essentially reduces it to a discrete sum.
\begin{table}
    \centering
    \begin{tabular}{|c|c|c|c|c|c|c|}
         \hline
         & $0$ & $1$ & $2$ & $3=\lfloor y_{critFut}^{n_1}\rfloor$ & ... & $n_1=10$\\
         \hline
        $0$ &  &  &  &  &  & \\
        \hline
        $1$ &  &  &  &  &  & \\
        \hline
        $2$ &  &  &  &  &  & \\
        \hline
        $3$ &  &  &  &  &  & \\
        \hline
        $...$ &  &  &  &  &  & \\
        \hline
        $22$ &  &  &  & x &  & \\
        \hline
        $23$ &  &  & x & x &  & \\
        \hline
        $24$ &  & x & x & x &  & \\
        \hline
        $25$ &  x & x & x & x &  & \\
        \hline
        $26$ &  x & x & x & x &  & \\
        \hline
        $27$ &  x & x & x & x &  & \\
        \hline
        $28$ &  x & x & x & x &  & \\
        \hline
        $29$ &  x & x & x & x &  & \\
        \hline
        $n_2-n_1=30$ & x & x & x & x &  & \\
        \hline
    \end{tabular}
    \caption{Illustration of \Cref{theorem:1}: Columns show the possible number of successes in the first batch of data $y^1=(y_1,...,y_{10})$ until the interim analysis is performed. Rows show the number of successes in the second batch of data $y^2=(y_{11},...,y_{30})$. The illustration uses $n_1=10$ and $n_2=40$, so after $n_1=10$ patients the interim analysis is carried out.}
    \label{tab:1}
\end{table}

Now, suppose we obtain for $n_1=10$ the critical value $y_{critFut}^{n_1}=3$ via root-finding. Suppose further that for $n_2=40$ we obtain $y_{critEff}^{n_2}=25$, also via root-finding as detailed in \cite{KelterPawel2025}. This implies that if we achieve $0,1,2$ or $3$ successes at the interim analysis, we would need $25,24,23$ or $22$ successes in the final analysis to sway the Bayes factor around to arrive at $\mathrm{BF}_{01}^{n_2}<k$. 

\Cref{tab:1} illustrates what \Cref{theorem:1} expresses: In the first sum, we iterate over the first $i=0$ to $\lfloor y_{critFut}^{n_1} \rfloor$ number of successes in the first batch of data for which $\mathrm{BF}_{01}^{n_1}>1/k$ holds, so the Bayes factor would stop for futility. In the example above, we used $\lfloor y_{critFut}^{n_1} \rfloor=3$ so we iterate from zero to three in the first sum, which are the columns of \Cref{tab:1}. Then, the first sum iterates over $y_i=0,...,3$ successes in this first batch of data. Now, if $y_s^1=0$, we have zero successes in the first batch of data and we need at least $25$ successes in the second batch of data $y^2=(y_{11},...,y_{40})$ to sway the Bayes factor around to yield $\mathrm{BF}_{01}^{n_2}(y^2)<k$ in the final analysis. Thus, for $y_s^1=0$ in the first column of \Cref{tab:1} the crosses mark the rows from $25$ to $30$ successes in the second batch of data, which are exactly the predictive probabilities we need to sum over for a fixed $y_s^1=0$ to yield $\mathrm{BF}_{01}^{n_2}(y^2)<k$ in the final analysis. The second sum in the right hand-side of \Cref{eq:theorem1} therefore sums over the prior-predictive probabilities $f(y_s^1=y_i,y_s^2=y_j|a_d,b_d,l,u)$ under $H_i$ for the values $y_j=\lceil y_{critEff}^{n_2}\rceil -y_i =25-y_i$ to $n_2-n_1=30$, and the first sum over $y_i=0$ to $y_i=\lfloor 3\rfloor$.

If $y_i=1$ (second column in \Cref{tab:1}) we obtain a single success in the first batch $y^1$ of data and the second sum in \Cref{eq:theorem1} iterates over $y_j=\lceil y_{critEff}^{n_2}\rceil -y_i =25-1=24$ to $n_2-n_1=30$ successes, so we obtain $25$ to $31$ successes in total then to sway the Bayes factor eventually around to $\mathrm{BF}_{01}^{n_2}(y^2)<k$ in the final analysis. Thus, the rows $24$ to $30$ are marked for column $1$ in \Cref{tab:1}.

\Cref{tab:1} thus shows graphically what \Cref{theorem:1} means: We iterate over the number of successes from $0$ to $\lfloor y_{critFut}^{n_1}\rfloor $, and for each value we iterate over the needed number of successes in the second batch of patient data of size $n_2-n_1$ to eventually sway the Bayes factor around in the final analysis, arriving at evidence in favour of $H_1$. This way, we compute the desired sum of marginal probabilities which equate to the probability $P(\mathrm{BF}_{01}^{n_2}<k|\mathrm{BF}_{01}^{n_1}>1/k,H_i)$.

\Cref{eq:theorem1} has one important advantage: It makes use only of the prior-predictive densities. As we do not condition on the observed data at interim, and take a prediction perspective which uses solely the prior before carrying out the trial, all elements in \Cref{eq:theorem1} can be computed easily. As shown in the proof of \Cref{theorem:1}, the prior-predictive can be computed via beta functions and regularized incomplete beta functions. The critical values $y_{critFut}^{n_1}$ and $y_{critEff}^{n_2}$ can be computed via the root-finding algorithm detailed in \cite{KelterPawel2025} numerically. Thus, \Cref{eq:theorem1} can be computed in a relatively straightforward manner.

Taking stock, this allows to compute power and type-I-error rates which adjust for the introduction of a single interim analysis in the two-stage design based on the trinomial tree branching.

\subsection{Computing the correct power for fixed sample sizes $n_1$ and $n_2$}
Now, the previous line of reasoning has led to the solution to obtain the correct power for a fixed sample size $n_2$, which accounts for the possibility to stop the trial for futility at the interim analysis. We must obtain the probability in \Cref{eq:partialPower} and subtract it from the power for sample size $n_2$, and we proceed in four steps:
\begin{enumerate}
    \item{Compute the unadjusted probability to stop due to futility at interim $P(\mathrm{BF}_{01}^{n_1}(y^1)>1/k\mid H_1)$ via root-finding based on $n_1$. The root-finding approach then yields a critical value $y_{critFut}^{n_1}$ for a fixed $n_1$.}
    \item{Compute the left-hand probability in \Cref{eq:conditional}, the unadjusted power $P(\mathrm{BF}_{01}^{n_2}(y) < k \mid H_1)$ at sample size $n_2$ -- for the final analysis -- via root-finding. The root-finding approach yields a critical value $y_{critEff}^{n_2}$ for a fixed $n_2$, based on which the Bayes factor would signal convincing evidence for $H_1$ at the end of the trial.}
    \item{Compute \Cref{eq:theorem1} for the values $y_{critFut}^{n_1}$ and $y_{critEff}^{n_2}$ obtained in the previous steps.}
    \item{Compute the \textit{futility-erased partial power} in \Cref{eq:partialPower} and subtract it from the unadjusted power $P(\mathrm{BF}_{01}^{n_2}(y) < k \mid H_1)$ computed in step two.}
\end{enumerate}  
Performing these four steps then yields an adjusted Bayesian power which accounts for the loss in power due to allowing to stop the trial early for futility.

\subsection{Computing the correct type-I-error rate for fixed sample sizes $n_1$ and $n_2$}
Computing the correct type-I-error rate for fixed sample sizes $n_1$ and $n_2$ proceeds likewise. The only difference is that \Cref{eq:partialPower} changes to
\begin{align}\label{eq:futilityAddedPartialTypeIError}
    \underbrace{P(\mathrm{BF}_{01}^{n_2}(y) < k \mid \mathrm{BF}_{01}^{n_1}(y^1)>1/k,H_0)}_{\text{Efficacy at final analysis given futility at interim}}\cdot \underbrace{P(\mathrm{BF}_{01}^{n_1}(y^1)>1/k\mid H_0)}_{\text{Futility at interim analysis}}
\end{align}
so $H_1$ is replaced by $H_0$. Essentially, we make use of \Cref{theorem:1} with $H_i=H_0$, so we perform the four steps in the last subsection with $H_1$ replaced by $H_0$. In step four, we could rephrase the \textit{futility-erased partial power} in \Cref{eq:conditional} as \textit{futility-erased partial type-I-error} then. Thus, \Cref{eq:futilityAddedPartialTypeIError} is called \textit{futility-erased partial type-I-error} henceforth.

\section{Calibrating the two-stage design}\label{sec:calibration}
Now, in this section we turn to the calibration of our design. The last section illustrated how to calculate the correct power and type-I-error rate of the Bayesian two-stage design when a single interim analysis is carried out. However, in these considerations, $n_1$ and $n_2$ were fixed. In practice, we do however specify the boundaries $\alpha$ and $1-\beta_{n_2}$ in \Cref{eq:bayesianDesign}, and want a calibration algorithm to reveal which sample sizes $n_1$ and $n_2$ we should use based on the specified boundaries for type-I-error rate and power. 

Two outcomes are possible: If we subtract the futility-erased partial power from the Bayesian power under $H_1$, our power may drop below $1-\beta_{n_2}$, compare \Cref{eq:bayesianDesign}. In that case, we must increase $n_2$ to account for the subtraction of the futility-erased partial power. 

On the other hand, under $H_0$, erasing the futility-erased partial type-I-error can only decrease the type-I-error rate. If two sample sizes $n_1$ and $n_2$ therefore are not calibrated with regard to the type-I-error rate in \Cref{eq:bayesianDesign}, it may happen that the design becomes fully calibrated once the futility-erased partial power is taken into consideration and subtracted from the original type-I-error rate obtained via root-finding for $n_2$.

\subsection{Calibration algorithm}\label{subsec:calibration}
The algorithm to calibrate the Bayesian two-stage design is therefore relatively simple:
\begin{enumerate}
    \item{Start with a sample size $n_2$ for which the requirement on Bayesian power in \Cref{eq:bayesianDesign} holds strictly. Such a sample size can be found via the root-finding approach outlined in \cite{PawelHeld2024} and \cite{KelterPawel2025}.}
    \item{Fix a smallest sample size $i\in \mathbb{N}$ and set $n_1:=i$ for the interim analysis. Iterate from $n_1:=i$ to $n_1:=n_2-1$. For each value, calculate the Bayesian power and type-I-error rates in \Cref{eq:bayesianDesign} and adjust these by subtracting the futility-erased partial power and futility-erased partial type-I-error. This step is crucial to obtain valid power and type-I-error rates, which account for multiple testing in the sequential design.}
    \item{If the Bayesian power drops below $1-\beta_{n_2}$ for the chosen $\beta_{n_2}$ (e.g. $1-\beta_{n_2}=0.80$) after reducing the power by the futility-erased partial power, or if the Bayesian type-I-error rate is larger than $\alpha$ for the chosen $\alpha$, or optionally, if the condition in \Cref{eq:futilityInterim} is not met, increase $n_1$ and repeat the last step. If $n_1=n_2-1$, increase $n_2$ and repeat the last step.}
\end{enumerate}
The later the interim analysis is carried out, that is, the larger $n_1$ is relative to $n_2$, the less probable it becomes that the Bayes factor signals evidence for $H_0$ when $H_1$ is true. As a consequence, the probability to stop early for futility and sway the Bayes factor around in the second stage becomes smaller and smaller for increasing $n_1$. Therefore, the futility-erased partial power decreases when increasing $n_1$. Increasing $n_2$ ensures that from the consistency of the Bayes factor, the power requirement in \Cref{eq:bayesianDesign} will eventually be met.\footnote{The algorithm thus increases $n_1$ only to the point until the futility-erased partial power is small enough to still yield a power larger than $1-\beta_{n_2}$}

Under $H_0$, the points above ensure that it becomes more and more difficult, and eventually impossible, to sway the Bayes factor around, when $n_1$ is increased. Thus, the futility-erased partial type-I-error decreases when increasing $n_1$. This ensures that increasing $n_1$ for a fixed $n_2$ calibrates the Bayesian type-I-error rate. If for the fixed $n_2$ and $n_1$ no solution can be obtained, increasing $n_2$ assures that the Bayesian power requirement will eventually be met, and increasing $n_1$ for that larger $n_2$ then calibrates the Bayesian type-I-error rate. Together, these aspects imply a calibration in terms of \Cref{eq:bayesianDesign}.

What about \Cref{eq:futilityInterim}? Under $H_0$, increasing $n_1$ for a fixed $n_2$ implies that the probability to stop early for futility at the interim analysis increases, so the condition
$$P(\mathrm{BF}_{01}^{n_1}(y^1)>1/k|H_0)>f$$
will eventually be met for large enough $n_1$, too. If no value $n_1$ can be found for which the latter holds, $n_2$ can be increased, thereby increasing the Bayesian power, decreasing the type-I-error rate, and allowing larger values of $n_1$ in turn. This ensures that eventually a large enough value $n_1$ is found for which the probability above passes the threshold $f$ under $H_0$, again due to the consistency of the posterior distribution \citep{Kleijn2022,VanDerVaart1998}. In some cases it might happen that the Bayesian power is larger than required, or the Bayesian type-I-error smaller than required, or even both, to satisfy the futility condition in the last display above. This is due to the discreteness of the binomially distributed data, a phenomenon which is well documented in the literature, see \cite{KelterPawel2025}. We return to this problem and how to solve it in the first example in the following section. In general, it remains difficult to say how the three design characteristics relate to another. We provide results for a variety of realistic settings in \Cref{sec:examples} below and return to this aspect.

\subsection{Possible issues with the calibration algorithm}
Suppose $n_1$ is quite small relative to the evidence threshold $k$ for the Bayes factor. Then, it might happen that for that $n_1$, there is no $y_{critFut}^{n_1}\in \{0,...,n_1\}$ so that for $y_1<y_{critFut}^{n_1}$ the condition $\mathrm{BF}_{01}^{n_1}(y^1)>1/k$ holds under the chosen design prior. As a consequence, the lowest blue branch in \Cref{fig:treePower} has probability zero, that is,
\begin{align}
    P(\mathrm{BF}_{01}^{n_1}(y^1)>1/k|H_i)=0
\end{align}
holds. In that case, the probability in \Cref{eq:futilityInterim} can exceed no threshold $f>0$ to stop for futility with at least probability $f$. As a consequence, the design cannot be calibrated for that choice of $n_1$, if the additional requirement in \Cref{eq:futilityInterim} is desired for the resulting sequential design, that is, if a minimum probability $f$ is warranted to obtain compelling evidence under $H_0$. Thus, the threshold $k$ based on which one stops for futility must be chosen not too conservative, so that there exists a minimal value for $n_1$ based on which the condition in \Cref{eq:futilityInterim} can hold. In all settings considered and reported in \Cref{sec:examples} below, this phenomenon never occurred. We found that setting the threshold at $k=3$ -- resembling moderate evidence -- always yields a calibrated design in our examples, while using the more strict $k=10$ -- resembling strong evidence -- is sometimes a too conservative choice and yields no solution to calibrate a design.

\subsection{The price of introducing an interim analysis}
Also, suppose that the conditions in \Cref{eq:bayesianDesign} hold for a value $n_2$. Then, the futility-erased partial power in \Cref{eq:partialPower} shows that the power of the two-stage design with an interim analysis is always smaller than the power of the design without an interim analysis under the following conditions:
\begin{theorem}\label{theorem:2}
    Let $n_1$ and $n_2$ be the interim analysis and final analysis sample sizes in the Bayesian two-stage design, $k$ be the evidence threshold and denote by $\mathrm{BF}_{01}^{n_1}(y^1)$ and $\mathrm{BF}_{01}^{n_2}(y)$ the Bayes factors for the interim and final analysis, where $y^1:=(y_1,...,y_{n_1})$ is the interim and $y:=(y_1,...,y_{n_1},y_{n_1+1},...,y_{n_2})$ the full trial data. If
\begin{align}\label{eq:theorem2}
    P(\mathrm{BF}_{01}^{n_2}(y) < k \mid \mathrm{BF}_{01}^{n_1}(y^1)>1/k,H_i)\cdot P(\mathrm{BF}_{01}^{n_1}(y^1)>1/k\mid H_i)\neq 0
\end{align}
holds for $i=1$, the Bayesian power in \Cref{eq:bayesianDesign} of the two-stage design for a fixed sample size $n_2$ is always smaller than the power of the same trial design without an interim analysis at sample size $n_1$, for any $n_1<n_2$. If \Cref{eq:theorem2} holds for $i=0$, the Bayesian type-I-error rate in \Cref{eq:bayesianDesign} of the two-stage design for a fixed sample size $n_2$ is always smaller than the Bayesian type-I-error rate of the same trial design without an interim analysis at sample size $n_1$, for any $n_1<n_2$.
\end{theorem}
\begin{proof}
    Follows from \Cref{eq:bayesianDesign}, \Cref{eq:conditional} and the composition of the Bayesian power as shown in \Cref{eq:adjustedBayesianPower} for the case of $i=1$, and from \Cref{eq:bayesianDesign}, \Cref{eq:conditional} and \Cref{eq:adjustedTypeIErrorRate} for the case of $i=0$.
\end{proof}
The result above clarifies the statistical price paid when introducing an interim analysis into a Bayesian design with Bayes factor testing. The power will almost always be smaller, while the Bayesian type-I-error rate will almost always improve. The condition for these facts to hold is, in other words, that the lowest blue branch in \Cref{fig:treePower} does not vanish by having probability zero.\footnote{If this is possible, because $n_1$ is too small to accumulate evidence $1/k$ in favour of $H_0$ via $\mathrm{BF}_{01}^{n_1}(y^1)>1/k$, the possibility of an interim analysis is not given anymore and the two-stage design reduces to a single-stage design.} Importantly, \Cref{theorem:2} provides a convenient option to check whether an optimal calibration as outlined in the next subsection is possible.

\subsection{Optimal calibration algorithm}\label{subsec:optimalCalibration}
We close this section by providing an optimal calibration algorithm in addition to the standard calibration algorithm detailed in \Cref{subsec:calibration}. In phase II trials, when $H_0$ holds, one would ideally stop at the interim analysis at sample size $n_1$ with probability one. On the other hand, if $H_1$ holds, and no stopping for efficacy is allowed -- which is a usual assumption in phase II trials, see \cite{Lee2008} -- it would be ideal if the trial always is carried out until $n_2$ patients are enrolled. Two popular strategies which go back to \cite{Simon1989} are to minimize the maximum number of patients which is enrolled, that is, $n_2$, or to minimize the expected number of patients under $H_0$, that is, $E[N|H_0]$. Expected sample size under $H_0$ is calculated as follows:
\begin{align}\label{eq:exp_sample_size_H0}
    E[N|H_0]=n_1 \cdot \underbrace{P(BF_{01}^{n_1}>1/k|H_0)}_{\text{stop the trial early for futility at $n_1$}}+ n_2 \cdot \underbrace{(1-P(BF_{01}^{n_1}>1/k|H_0))}_{\text{continue the trial until $n_2$}}
\end{align}
Expected sample size under $H_1$ is calculated likewise as follows:
$$E[N|H_1]=n_1 \cdot \underbrace{P(BF_{01}^{n_1}>1/k|H_1)}_{\text{stop the trial early for futility at $n_1$}}+ n_2 \cdot \underbrace{(1-P(BF_{01}^{n_1}>1/k|H_1))}_{\text{continue the trial until $n_2$}}$$

When minimizing $E[N|H_0]$, \Cref{eq:exp_sample_size_H0} shows that this implies that the probability to stop for futility under $H_0$ is maximized, while the probability to continue the trial until $n_2$ under $H_0$ is minimized. This is the approach taken in two-stage phase II optimal design of \cite{Simon1989}, and we provide an analogue optimal Bayesian sequential Bayes factor design by minimizing $E[N|H_0]$ under the constraints that \Cref{eq:bayesianDesign} holds.
\begin{definition}[Optimal two-stage Bayes factor design]
    Let $\alpha \in (0,1)$ and $\beta_{n_2}\in (0,1)$ be given and denote by $n_{min}$ the minimum sample size at which the trial is allowed to stop for futility and by $n_{max}$ the maximum trial size of the final analysis. The optimal two-stage Bayes factor design with interim sample size $n_1$ and full sample size $n_2$, is the design for which $n_1$ and $n_2$ are the solution to the minimization problem
    \begin{equation}
        \begin{aligned}
            \min_{n_1,n_2} \quad & E[N|H_0]\\
            \textrm{s.t.} \quad & P(\mathrm{BF}_{01}^{n_2}(y) < k \mid H_0) \leq \alpha \hspace{1cm}\\
            & \text{and} \hspace{1cm} P(\mathrm{BF}_{01}^{n_2}(y) < k \mid H_1) \geq 1-\beta_{n_2}\\
            & \text{and} \hspace{1cm} n_{min}\leq n_1 < n_2\leq n_{max}    \\
        \end{aligned}
    \end{equation}
\end{definition}
Optionally, we can provide an optimal Bayesian sequential Bayes factor design by additionally require \Cref{eq:futilityInterim} to hold. The optimal calibration algorithm is relatively simple: Provide a minimum value $n_{min}$ for $n_1$ and a maximum value $n_{max}$ for $n_2$ based on subject-domain knowledge about how many patients can realistically be enrolled in a given timeframe. Often, this sets an upper bound $n_{max}$ on $n_2$. The minimum value $n_{min}$ for $n_1$ can be set as small as wanted, e.g. $n_1=4$ or $n_1=5$. In this range, compute the design characteristics in \Cref{eq:bayesianDesign} (and optionally in \Cref{eq:futilityInterim}), and filter those values of $n_1$ and $n_2$ which are calibrated by running the calibration algorithm in \Cref{subsec:calibration}. Next, compute $E[N|H_0]$ for these calibrated designs and return the design which minimizes the latter.

Before turning to the examples, we return briefly to \Cref{theorem:2}. Based on \Cref{theorem:2}, we know that a sequential design's power can only decrease compared to the power the same design without an interim analysis. For that same design without an interim analysis, we can compute the power and type-I-error rate almost instantaneously via root-finding \citep{KelterPawel2025}. Thus, when a value $n_{max}$ is provided, we can first check if the power of the non-sequential design is above $1-\beta_{n_2}$. If not, \Cref{theorem:2} assures that the two-stage design cannot be calibrated and we must increase $n_{max}$. This can speed up the calibration algorithm even further.

\section{Examples}\label{sec:examples}
In this section, we provide the results of the sequential two-stage design for a variety of practically relevant settings for a phase-II-a trial with binary endpoint. We stress that the results here are not simulation results but can be obtained in only seconds by applying the calibration algorithm outlined in \Cref{sec:calibration}. Also, no Monte Carlo standard error is needed.

Regarding the relevant design characteristics, we provide the (Bayesian and frequentist) power, expected sample size under $H_0$, type-I-error rates (Bayesian and frequentist) and the probability to obtain compelling evidence under $H_0$.

\subsection{An example showing why oscillations cause problems}\label{subsec:ex_oscillations}
We start with an example which returns to the fact that oscillations are a well-known phenomenon for the beta-binomial model. Thus, we test $H_0:p\leq 0.1$ against $H_1:p>0.1$, and aim for $\alpha=0.05$ and $\beta_{n_2}=0.2$, so we cap the type-I-error rate at 5\% and require at least 80\% power. We do not add the requirement of a minimum probability $f$ for compelling evidence under $H_0$ right now. For the optimal Bayesian designs, we search for sample sizes which minimize $E[N|H_0]$ in the range of $n_1=5$ to $n_2=40$, so $n_{min}=5$ and $n_{max}=40$. Also, we always use flat analysis priors in the final analysis, while we investigate the design's operating characteristics under several design priors used during the planning stage of the trial.

\begin{table}[h!]
    \centering
    \resizebox{\textwidth}{!}{\begin{tabular}{lcccccc}
    \hline
  Design & $n_1$ & $n_2$ & type-I-error & power & $E[N|H_0]$ & PET($p_0$)\\
    \hline
  Simon's 2-stage phase II design &  &  &  &  &  &  \\
  \hspace{0.25cm} minimax design & 15	& 25 & 0.0328 & 0.8017 & 19.51 & 0.5490\\
  \hspace{0.25cm} optimal design & 10	& 29 & 0.0471 & 0.8051 & 15.01 & 0.7361\\
  \hline
  Optimal sequential Bayes factor design &  &  &  &  &  & PCE($p_0$)\\
  \hspace{0.25cm} frequentist power, $k=1/3$, $k_f=3$ & 10 & 29 & 0.0471 & 0.8051 & 15.01 & 0.7361\\
  \hspace{0.25cm} Bayesian power, $k=1/3$, $k_f=3$ & & & & & & \\
  \hspace{1cm} with $\mathrm{Beta}_{[0.1,1]}(1,1)$ prior & 5 & 15 & 0.0480 & 0.8107 & 9.10 & 0.5905\\
  \hspace{1cm} with $\mathrm{Beta}_{[0.1,1]}(7,15)$ prior & 11 & 36 & 0.0470 & 0.8017 & 18.57 & 0.6974\\
  \hspace{1cm} with $\mathrm{Beta}_{[0.1,1]}(11.29,25)$ prior & 12 & 28 & 0.0477 & 0.8020 & 17.46 & 0.6590\\
  \hspace{1cm} with $\mathrm{Beta}_{[0.1,1]}(22,50)$ prior & 10 & 36 & 0.0432 & 0.8038 & 16.86 & 0.7361\\
    \hline
    \end{tabular}}
    \caption{Design characteristics for Simon's two-stage design and different Bayesian sequential designs for Example 1. $n_1$ and $n_2$ denote the interim and full sample sizes, the type-I-error is always computed via a point prior at $p_0$ for all Bayesian designs, while the power is computed as frequentist power with a point prior at $p_1=0.3$, or as Bayesian power with a truncated $\mathrm{Beta}_{[0.1,1]}(a_d,b_d)$ prior, where $a_d=b_d=1$ denotes the flat prior. Larger values of $a_d$ and $b_d$ imply a more informative prior, see \Cref{fig:centered_beta_priors_p0.3}. Bayesian designs are calibrated as optimal so that they minimize the expected sample size $E[N|H_0]$ under $H_0$. $k$ denotes the threshold used in Bayes factor designs for type-I-error rates and power in \Cref{eq:bayesianDesign}, and $k_f$ denotes the threshold used for the additional requirement for compelling evidence under $H_0$ in \Cref{eq:futilityInterim}. PET($p_0$) denotes the probability of early termination for Simon's designs, which is the probability to stop the trial early for futility. PCE($p_0$) denotes the probability of compelling evidence in favour of $H_0$, which is computed with a point prior at $p_0$, and based on the Bayes factor threshold $k_f$ for futility.}
    \label{tab:ex_oscillations}
\end{table}
\begin{figure}[h!]
    \centering
    \includegraphics[width=1\linewidth]{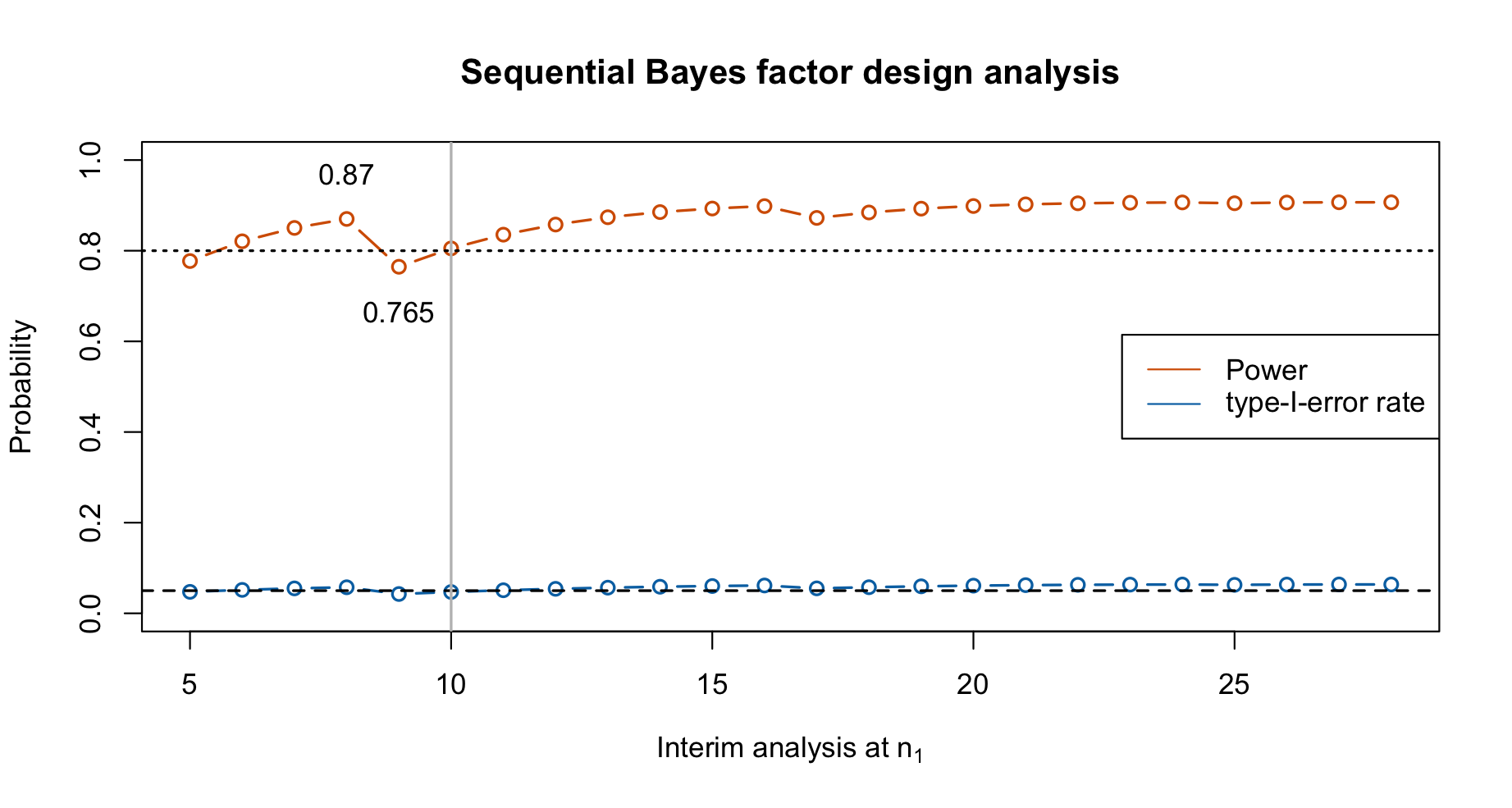}
    \caption{Design characteristics for the first example, showing how oscillations occur due to the discreteness of the binomially distributed data. Vertical line shows the value $n_1=10$ which is the optimal value that, together with $n_2=29$ yields a calibrated sequential Bayes factor design based on frequentist power computed at $p_1=0.3$, and which is optimal in the sense that it minimizes the expected sample size $E[N|H_0]$.}
    \label{fig:ex_oscillations}
\end{figure}
\Cref{tab:ex_oscillations} shows the results for Simon's optimal and minimax designs\footnote{These can either be computed via the online calculator of the UNC Lineberger Comprehensive Cancer Center at \url{http://www.cancer.unc.edu/biostatistics/program/ivanova/SimonsTwoStageDesign.aspx} or via the R software package \texttt{clinfun}, available at \url{https://cran.r-project.org/web/packages/clinfun/index.html}.} and for the optimal sequential Bayes factor design using frequentist or Bayesian power and the evidence thresholds $k=1/3$ and $k_f=3$ for evidence and futility. Thus, at the interim analysis we check whether $\mathrm{BF}_{01}^{n_1}(y^1)>k_f$ to stop for futility, and at the final analysis we check $\mathrm{BF}_{01}^{n_2}(y)<k$ to test whether $H_1:p>p_0$ at the full sample size $n_2$ at the end of the trial holds. Note that the Bayesian design calculates the type-I-error rate strictly frequentist using a point prior at $p_0=0.1$, while Bayesian power is also strictly frequentist, using a point prior at $p_1=0.3$. Thus, the Bayesian design has a fully frequentist interpretation here. \Cref{fig:ex_oscillations} shows how oscillations occur for different values of $n_1$. Thus, it might happen that the power drops below the desired thresholf of $80\%$, e.g. from $0.87$ at $n_1=8$ to $0.765$ at $n_1=9$ in \Cref{fig:ex_oscillations}. A possible solution to this phenomenon is to require the power to stay above the threshold $\beta_{n_2}$ (and likewise, to require the type-I-error rate to stay below $\alpha$) for the next $10$ natural numbers, see also the discussion in \cite{KelterPawel2025}. This is the strategy adopted here and which in all examples we carried out yielded reliable results.\footnote{For details, see the \texttt{bfpwr} package on CRAN: \url{https://cran.r-project.org/web/packages/bfpwr/index.html}.} We recommend plotting the design's power and type-I-error rate for all interim analyses between $n_{min}$ and $n_{max}$ like in \Cref{fig:ex_oscillations} to check whether the resulting power remains calibrated.\footnote{This is also relevant in case a trial is stopped due to too many severe adverse events. Then, recalculation of the power and type-I-error rate is possible by plots such as \Cref{fig:ex_oscillations} even when a deviation from the trial protocol became necessary.}

\Cref{tab:ex_oscillations} provides several insights: First, the optimal sequential Bayes factor design with frequentist power and $k=1/3$ and $k_f=3$ recovers Simon's two-stage optimal design. All design characteristics are identical, although the concept of probability of compelling evidence under $H_0$ -- that is, PCE($p_0$) in \Cref{tab:ex_oscillations} -- which is computed via a point prior at $p_0=0.1$, is stronger than the concept of probability of early termination under $H_0$ -- that is, PET($p_0$) in \Cref{tab:ex_oscillations}. PCE($p_0$) is the probability of compelling evidence under $H_0$ and implies that we actually have moderate evidence in favour of $H_0:p\leq 0.1$, while PET($p_0$) only implies that we have not sufficient evidence in favour of $H_1:p>0.1$. Absence of evidence is, however, no evidence of absence \citep{Altman1995}.

Second, shifting to Bayesian power and inspecting the optimal sequential Bayes factor design characteristics demonstrates the influence of the truncated Beta prior on $H_1$ to calculate the power. For a flat $\mathrm{Beta}_{[0.1,1]}(1,1)$ prior, larger probabilities like $p=0.5$ or $p=0.8$ have the same a priori probability than smaller ones such as $p=0.25$. As a consequence, the sample sizes $n_1$ and $n_2$ for which a calibrated design can be found become smaller. This is, in particular, because the choice of $n_1$ and $n_2$ is primarily driven by the requirement on power, compare \Cref{fig:ex_oscillations}. \Cref{tab:ex_oscillations} shows that the sample sizes decrease to $n_1=5$ and $n_2=15$, and the design is still calibrated under a Bayesian notion of power, while the type-I-error-rate is still calibrated from a frequentist point of view (that is, under a point prior on $p_0=0.1$). Due to the smaller values $n_1=5$ and $n_2=15$ compared to $n_1=10$ and $n_2=29$ under a strictly frequentist power on $p_1=0.3$, the resulting expected sample size under $H_0$ (and $H_1$) reduces, too. Therefore, the expected sample size under $H_0$ beats the one of Simon's two-stage optimal design above, but the price is the shift to a Bayesian concept of power. Also, the probability of compelling evidence under $H_0$ decreases a little, because $n_1$ is smaller now and it becomes more difficult for the Bayes factor to accumulate moderate evidence in favour of $H_0$ after only $n_1=5$ instead of $n_1=10$ patients.

\begin{figure}[h!]
    \centering
    \includegraphics[width=1\linewidth]{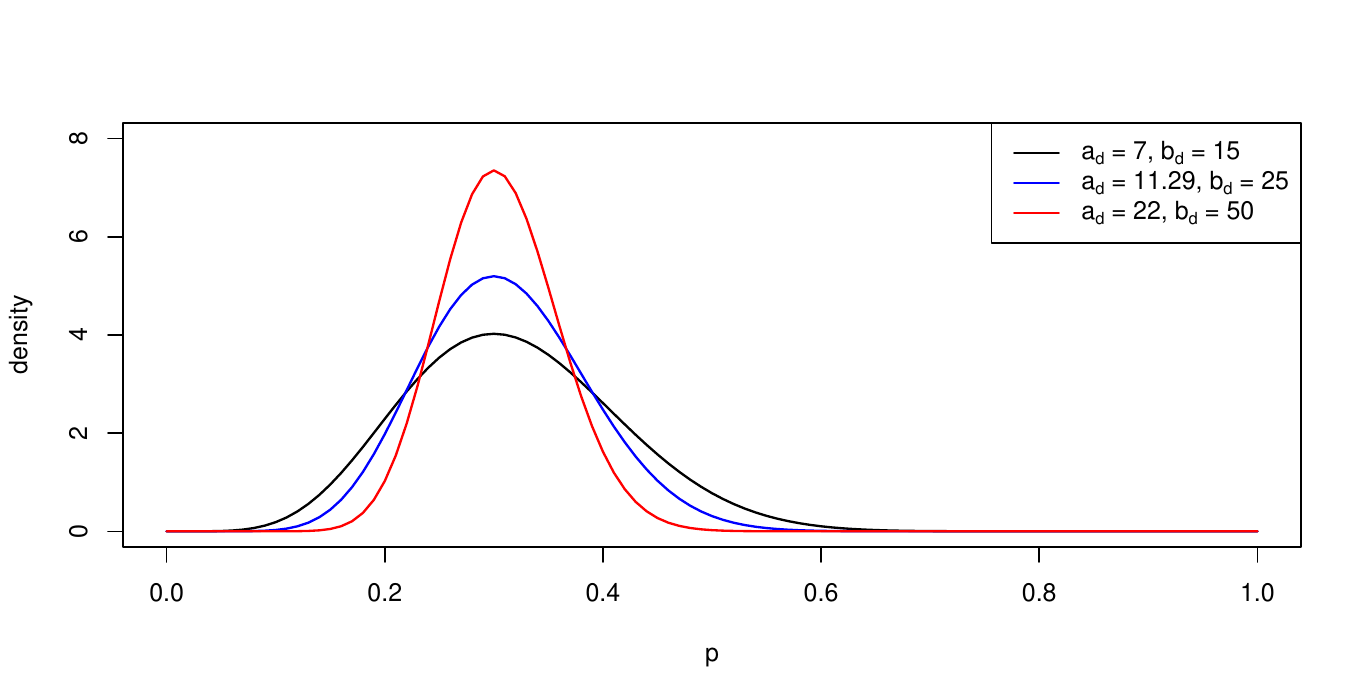}
    \caption{Informative $\mathrm{Beta}_{[0.1,1]}(a_d,b_d)$ priors used in the first example. A prior with larger sum of $a_d+b_d$ corresponds to a more informative prior, where $a_d$ can be interpreted as already observed failures and $b_d$ as already observed successes.}
    \label{fig:centered_beta_priors_p0.3}
\end{figure}

Note that a flat prior on $[0.1,1]$ might be unrealistic, because extremely large probabilities such as $p=0.9$ are thought of equally probable a priori as small ones like $p=0.15$. As a consequence, it might be more realistic to choose a prior centered at the frequentist point prior value $p_1=0.3$. We proceed to this option now and investigate how the design characteristics change. \Cref{fig:centered_beta_priors_p0.3} shows Beta priors centered at $p=0.3$ for increasing informativity, and \Cref{tab:ex_oscillations} includes the optimal sequential Bayes factor design characteristics for these three priors. Two aspects are noteworthy: First, the more informative the prior, the smaller the expected sample size becomes. This is to be expected, because the prior concentrates more closely around $p=0.3$ and puts less probability mass on smaller values close to $p_0=0.1$. Second, the PCE($p_0$) is not influenced by the prior on $H_1$, and differences are solely attributed to the changes in $n_1$ and $n_2$.\footnote{As a sidenote, when adding the requirement of $\mathrm{PCE}(p_0)>f$ for $f=0.7$ to the sequential Bayes factor design with the flat Beta prior on $H_1$, it also recovers Simon's optimal two-stage design as a special case in this example.}
\subsection{Single-arm phase II trial for nonsmall cell lung cancer}\label{subsec:phaseII}

In this subsection, we adopt an example of a clinical trial discussed in \cite{KelterSchnurr2024}. A single-arm phase IIA study to demonstrate the efficacy of a novel combination therapy as front-line
treatment in patients with advanced nonsmall cell lung cancer is considered and the primary endpoint is binary. The primary objective of the study was to assess the efficacy of the novel treatment, which involved the combination of a vascular endothelial growth factor antibody plus an epidermal growth factor receptor tyrosine kinase inhibitor. The primary endpoint is the clinical
response rate, that is, the rate of complete response and partial response combined, to the new treatment. The hypotheses of interest are 
$$H_0:p\leq p_0 \text{ versus } H_1:p>p_1$$
The current standard treatment yields a response rate of $\approx$ 20\%, so that $p_0=0.2$. The target response rate of the new regimen is 40\%, so $p_1 = 0.4$. We stress that somewhat unrealistic, the parameter interval $(0.2,0.4]$ is somehow excluded from the hypotheses considered, a choice sometimes seen in phase IIA studies, compare \cite{Lee2008}. While such a distance between $H_0$ and $H_1$ allows for an easier separation between both hypotheses via statistical testing,  we refrain from doing so. We provide the optimal sequential Bayes factor design under truncated Beta design priors centered on $p_1=0.4$ with varying informativity, so we select $0.4$ for the mode of the design prior and opt for a slightly informative $\mathrm{Beta}_{[0.2,1]}(10.33,15)$ prior, and a strongly informative $\mathrm{Beta}_{[0.2,1]}(13.66,20)$ prior. Also, we provide results under a flat Beta prior, and using frequentist power. Like always, type-I-error rates are calculated as frequentist error rates with a point prior at $p_0$. For the analysis priors, we select flat truncated Beta priors both under $H_0$ and $H_1$, which is the usual choice.

Now, in \cite{KelterSchnurr2024}, a sequential design was used and calibrated via a Monte Carlo simulation to attain the boundaries $\alpha\leq 0.1$ and $\beta \leq 0.1$ for the Bayesian type-I-error rate and power. With the novel idea of trinomial tree branching and the derived results, no simulation is necessary to compute the optimal sequential Bayes factor design now. We also adopt the choice $\alpha\leq 0.1$ and $\beta \leq 0.1$. Now, instead of conducting Monte Carlo simulations, we use trinomial tree branching, correct the required probabilities, and carry out our calibration algorithm to yield a design for which
$P(\mathrm{BF}_{01}^{n_2}(y) < k \mid H_0) \leq \alpha$ and $P(\mathrm{BF}_{01}^{n_2}(y) < k \mid H_1) \geq 1-\beta_{n_2}$ for $\alpha = 0.1$ and $\beta_{n_2}=0.2$, compare \Cref{eq:bayesianDesign}.  

In addition, we add the requirement $P(\mathrm{BF}_{01}^{n_1}(y^1)>k_f|H_0)>f$, compare \Cref{eq:futilityInterim}, in some cases in the following example. Thus, we can add the requirement that we obtain compelling evidence under $H_0$, e.g. for $f=0.6$ we have 60\% probability to reach a Bayes factor that passes our threshold for evidence. For notational clarity, we denote this threshold as $k_f$ (for futility), because we could choose a different value $k_f$ to stop for futility at $n_1$ than to stop for evidence based on threshold $k$ at $n_2$. To obtain an optimal sequential Bayes factor design with out approach, we analyzed the metrics of the resulting designs in the range of $n_{min}=5$ to $n_{max}=60$, while for some cases (with very informative priors and high requirements on power, e.g. 90\%) we had to increase $n_{max}$ up to $n_{max}=150$ in the most extreme case. However, this happened solely under the strongly informative prior and a evidence threshold for strong evidence according to \cite{Jeffreys1939}.

\begin{table}[h!]
    \centering
    \begin{tabular}{lccccccc}
    \hline
  Design & $n_1$ & $n_2$ & type-I-error & power & $E[N|H_0]$ & PET($p_0$)\\
    \hline
  Simon's 2-stage phase II &  &  &  &  &  &  & \\
  \hspace{0.25cm} minimax design & 19	& 36 & 0.0861 & 0.9024 & 28.26	& 0.4551\\
  \hspace{0.25cm} optimal design & 17	& 37 & 0.0948 & 0.9033 & 26.02 & 0.5489\\
  \hline
  Optimal sequential Bayes factor &  &  &  &  &  & PCE($p_0$)\\
  \hspace{0.25cm} frequentist power & & & & & & & \\
  \hspace{1cm}  $k=1/3$, $k_f=3$ & 17 & 37 & 0.0948 & 0.9033 & 26.02 & 0.5489 & \\
  \hspace{1cm}  $k=1/3$, $k_f=3$, $f=0.6$ & 30 & 36 & 0.0886 & 0.9091 & 32.36 & 0.6070 & \\
  \hspace{1cm}  $k=1/10$, $k_f=3$ & 21 & 51 & 0.0340 & 0.9021 & 33.42 & 0.5860 & \\
  \hspace{1cm}  $k=1/10$, $k_f=3$, $f=0.6$ & 30 & 50 & 0.0302 & 0.9011 & 37.86 & 0.6070 & \\
  \hspace{0.25cm} Bayesian power & & & & & & & \\
  \hspace{0.5cm} $k=1/3$, $k_f=3$ & & & & & & & \\
  \hspace{1cm}  $\mathrm{Beta}_{[0.2,1]}(1,1)$ prior & 27 & 54 & 0.0988 & 0.9003 & 39.46 & 0.5387\\
  \hspace{1cm}  $\mathrm{Beta}_{[0.2,1]}(10.33,15)$ prior & 28 & 67 & 0.0981 & 0.9006 & 47.48 & 0.5005\\
  \hspace{1cm}  $\mathrm{Beta}_{[0.2,1]}(13.66,20)$ prior & 24 & 58 & 0.0958 & 0.9001 & 42.36 & 0.4599\\
  \hspace{0.5cm} $k=1/10$, $k_f=3$ & & & & & & & \\
  \hspace{1cm}  $\mathrm{Beta}_{[0.2,1]}(1,1)$ prior & 42 & 100 &  0.0325 & 0.9002 & 69.21 & 0.5309\\
  \hspace{1cm}  $\mathrm{Beta}_{[0.2,1]}(10.33,15)$ prior & 65 & 100 & 0.0339 & 0.9001 & 79.93 & 0.5735\\
  \hspace{1cm}  $\mathrm{Beta}_{[0.2,1]}(13.66,20)$ prior & 40 & 88 & 0.0347 & 0.9005 & 59.53 & 0.5931\\
  \hspace{0.5cm} $k=1/3$, $k_f=3$, $f=0.6$ & & & & & & & \\
  \hspace{1cm}  $\mathrm{Beta}_{[0.2,1]}(1,1)$ prior & 30 & 58 & 0.0928 & 0.9008 & 41.00 & 0.6070\\
  \hspace{1cm}  $\mathrm{Beta}_{[0.2,1]}(10.33,15)$ prior & 30 & 76 & 0.0923 & 0.9008 & 48.08 & 0.6070\\
  \hspace{1cm}  $\mathrm{Beta}_{[0.2,1]}(13.66,20)$ prior & 30 & 63 & 0.0978 & 0.9056 & 42.97 & 0.6070\\
  \hspace{0.5cm} $k=1/10$, $k_f=3$, $f=0.6$ & & & & & & & \\
  \hspace{1cm}  $\mathrm{Beta}_{[0.2,1]}(1,1)$ prior & 30 & 130 & 0.0279 & 0.9016 & 69.30 & 0.6070\\
  \hspace{1cm}  $\mathrm{Beta}_{[0.2,1]}(10.33,15)$ prior & 30 & 147 & 0.0263 & 0.9011 & 75.98 & 0.6070\\
  \hspace{1cm}  $\mathrm{Beta}_{[0.2,1]}(13.66,20)$ prior & 30 & 108 & 0.0273 & 0.9001 & 60.66 & 0.6070\\
    \hline
    \end{tabular}
    \caption{Design characteristics for Simon's two-stage design and different Bayesian sequential designs for Example 1. $n_1$ and $n_2$ denote the interim and full sample sizes, the type-I-error is always computed via a point prior at $p_0$ for all Bayesian designs, while the power is computed as frequentist power with a point prior at $p_1=0.4$, or as Bayesian power with a truncated $\mathrm{Beta}_{[0.2,1]}(a_d,b_d)$ prior, where $a_d=b_d=1$ denotes the flat prior. Larger values of $a_d$ and $b_d$ imply a more informative prior, see \Cref{fig:centered_beta_priors_p0.3}. Bayesian designs are calibrated as optimal so that they minimize the expected sample size $E[N|H_0]$ under $H_0$. $k$ denotes the threshold used in Bayes factor designs for type-I-error rates and power in \Cref{eq:bayesianDesign}, and $k_f$ denotes the threshold used for the additional requirement for compelling evidence under $H_0$ in \Cref{eq:futilityInterim}. PET($p_0$) denotes the probability of early termination for Simon's designs, which is the probability to stop the trial early for futility. PCE($p_0$) denotes the probability of compelling evidence in favour of $H_0$, which is computed with a point prior at $p_0$, and based on the Bayes factor threshold $k_f$ for futility.}
    \label{tab:ex2}
\end{table}
\Cref{tab:ex2} on the next page provides the resulting characteristics of the different designs: The optimal calibration algorithm for the sequential Bayes factor design recovers Simon's optimal design, compare the values for the Bayesian design with frequentist power, $k=1/3$, $k_f=3$. Thus, in this example, under a moderate threshold $k$ for evidence and for futility $k_f$, Simon's optimal design is a special case of our Bayesian design when power is computed under a frequentist point prior.

Second, when adding the requirement of compelling evidence to the optimal sequential Bayes factor designs, the sample size of the interim analysis often increases. For example, for Bayesian power with $k=1/3$ and $k_f=3$, the sample sizes increase from $n_1=27$, $n_1=28$ and $n_1=24$ to $n_1=30$ when adding $f=0.6$ as the minimum probability for $\mathrm{PCE}(p_0)$. One possible reason is that to accumulate 60\% probability to reach a Bayes factor of at least $3$ in favour of $H_0$, the interim sample size $n_1$ must be large enough to express that much evidence statistically. When adding such a requirement, the expected sample size thus increases slightly, as a consequence. Still, there is no option to add this additional layer of evidence on e.g. Simon's optimal design. When the trial is stopped there for futility, there is not sufficient evidence to reject $H_0$. However, absence of evidence is no evidence of absence, so it remains unclear whether $H_0$ holds true in that case.\footnote{Likewise, a large p-value close to $1$ does not imply that $H_0$ is true.} For the Bayesian design with that additional requirement, once the trial is stopped at $n_1=30$, we can be sure that if $H_0$ is true, we arrive with 60\% at a Bayes factor $\mathrm{BF}_{01}>k_f=3$. Thus, we have a 60\% probability to find moderate evidence for $H_0$ during the interim analysis, if $H_0$ holds. The result $\mathrm{BF}_{01}>k_f=3$ implies at least moderate evidence in favour of $H_0$, and absence of evidence becomes presence of evidence (for $H_0$). We can be moderately certain that $H_0$ is true when we stop for futility for such a design. If we had chosen $k_f=10$, we could even be strongly certain. This highlights an important aspect of the proposed design: Speaking of stopping early for futility is not appropriate. More appropriate is to speak of stopping for compelling evidence in favour of the null hypothesis, which shows that it is possible to accept $H_0$ even after the interim analysis at $n_1$.

Third, \Cref{tab:ex2} shows that frequentist power calculations could be regarded as much too optimistic, in particular, when inspecting the result under different Beta priors for the Bayesian designs. It remains entirely unclear how effective the drug is, and indeed, efficacy probabilities in the range between $p=0.2$ and $p=0.4$ are possibly very likely. Frequentist power was computed under $p_1=0.4$, however. As a consequence Bayesian power with $k=1/3$ and $k_f=3$ could be more realistic, because a flat Beta design prior distributes the probability mass under $H_1:p>0.2$ evenly. Thus, when shifting from frequentist power to Bayesian power with $k=1/3$ and $k_f=3$, and a flat $\mathrm{Beta}_{[0.2,1]}(1,1)$ prior, the sample sizes $n_1$ and $n_2$ increase to $n_1=27$ and $n_2=54$, and the expected sample size under $H_0$ increases to $39.46$. Adding the requirement of compelling evidence under $H_0$ with 60\% on top, these sample sizes increase even slightly further. Now, a flat prior could be criticised as a very large probability such as $p=0.9$ gets the same prior probability as a small one like $p=0.25$, but in most cases one would be quite uncertain about high efficacy probabilities in a phase II study. Thus, the Beta priors centered at $p_1=0.4$ could be more realistic. The informativity of the $\mathrm{Beta}_{[0.2,1]}(10.33,15)$ prior is equal to having observed $10.33$ successes and $15$ failures already \citep{Kruschke2018}, which is not too much for a phase II study. The $\mathrm{Beta}_{[0.2,1]}(13.66,20)$ prior is more informative, and implies that $\approx 14$ failures and $20$ successes have already been recorded. The bulk of prior probability mass locates therefore around $p_1=0.4$, which is the mode of both priors. \Cref{tab:ex2} shows that for such more realistic priors, the sample sizes $n_1$ and $n_2$ as well as $E[N|H_0]$ are, in general, slightly larger compared to frequentist power calculations. Two aspects are worth mentioning: 
\begin{enumerate}
    \item[$\blacktriangleright$]{The results in \Cref{tab:ex2} show that shifting from a flat prior to a slightly informative one does not necessarily reduce the required sample sizes. This is due to the fact, that probabilities close to $1$ are downweighed strongly even for slightly informative priors, compare \Cref{fig:centered_beta_priors_p0.3}. Only if a certain amount of informativity is reached, small success probabilities close to $p_0$ are downweighed enough so that the sample size starts decreasing again. For example, for $k=1/3$ and $k_f=3$, $E[N|H_0]$ progresses from $39.46$ for the flat prior, over $47.48$ for the slightly informative prior, to $42.36$ for the strongly informative prior. A similar pattern is visible for $k=1/10$ and $k_f=3$ and all other scenarios.}
    \item[$\blacktriangleright$]{Shifting from $k=1/3$ to $k=1/10$ yields a much larger $n_2$ and a larger $n_1$ in most cases, because more observations are necessary for the Bayes factor to express strong evidence compared to only moderate evidence. As a consequence, for different thresholds $k\neq 1/k_f$ such as $k=1/10$ and $k_f=3$, the type-I-error rates and power become imbalanced in the sense that the type-I-error rate is not exhausting the 10\% upper limit anymore (it is more close to $\approx 3\%$ for $k=1/10$. Thus, the sample size is driven primarily by the power requirement in these cases. We therefore recommend similar choices for $k$ and $k_f$ to exhaust the design's restrictions in \Cref{eq:bayesianDesign}.}
\end{enumerate}
In summary, the results shown in \Cref{tab:ex2} show how flexible the optimal sequential Bayes factor design is. It provides a fully Bayesian interpretation to Simon's two-stage optimal design. Still, if a more flexible power calculation with the option to incorporate prior information is warranted, shifting to Bayesian notions of power might be helpful. The price is an increased sample size at the benefit of more realism. Adding a threshold for a minimum probability for compelling evidence under $H_0$ comes at a price, too, but adds a layer of evidence to the analysis, in particular, when the trial is stopped at $n_1$.

As a last note, the second Bayesian design with frequentist power and $k=1/3$, $k_f=3$ and $f=0.6$ in \Cref{tab:ex2} can be interpreted as a generalized Simon's optimal design with a Bayesian validity: It provides frequentist type-I-error rates and power, but a Bayesian notion of compelling evidence for $H_0$. Thus, it is like a Simon's optimal design where additionally the probability threshold to find compelling evidence under $H_0$ can be specified.

\subsection{Comparison with non-sequential Bayes factor design analysis}
\begin{figure}[!h]
    \centering
    
     \begin{subfigure}[b]{1\textwidth}
         \centering
         \includegraphics[width=\textwidth]{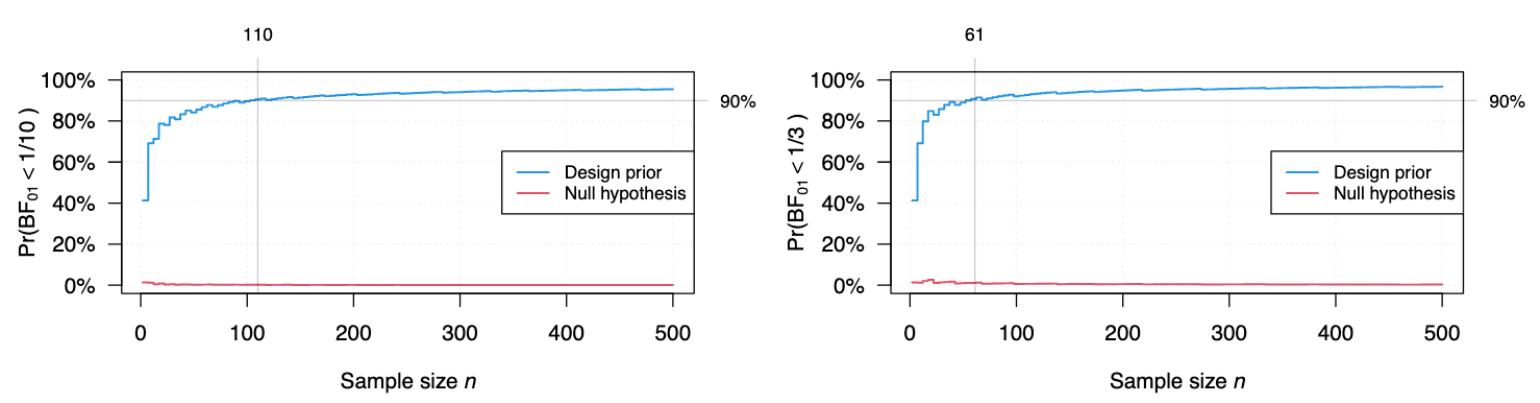}
         \caption{Evidence threshold $k=1/10$ (left) and $k=1/3$ right}
         \label{fig:ex3_1}
     \end{subfigure}
     \hfill
     \begin{subfigure}[b]{1\textwidth}
         \centering
         \includegraphics[width=\textwidth]{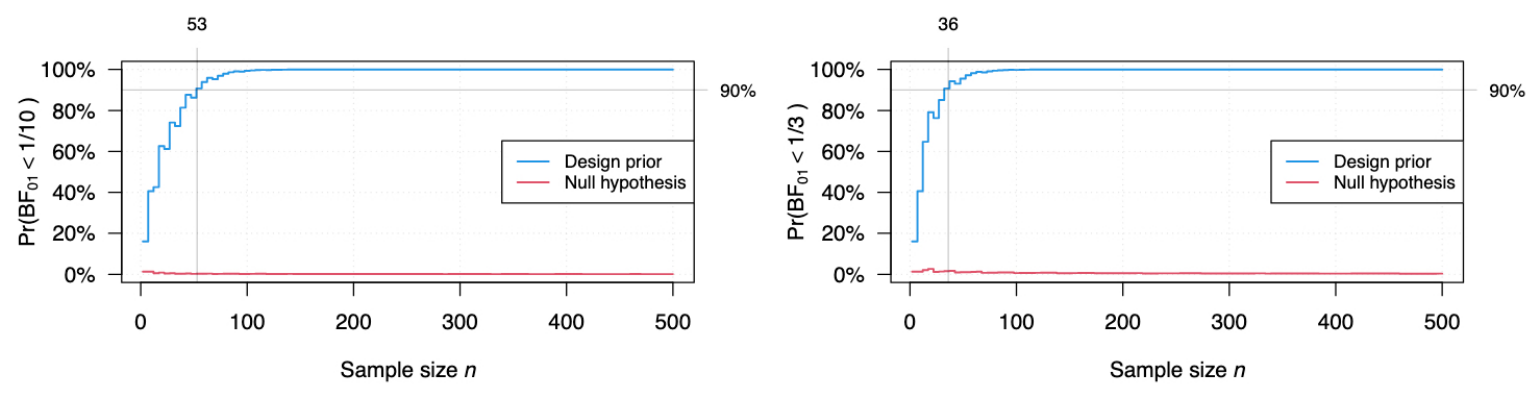}
         \caption{Evidence threshold $k=1/10$ (left) and $k=1/3$ (right)}
         \label{fig:ex3_2}
     \end{subfigure}
     
    \caption{Bayesian power and type-I-error rate for the single-arm phase II proof of concept trial with binary endpoint for $k=1/3$ (left) and $k=1/10$ (right). Top plots show the Bayesian power $P(\mathrm{BF}_{01}(y)<k \mid H_1)$ under $H_1$ as the solid blue line, and the Bayesian type-I-error rate $P(\mathrm{BF}_{01}(y)<k \mid H_0)$ under $H_0$ as the solid red line. A flat $\mathrm{Beta}_{[0.2,1]}(1,1)$ prior was used. The bottom plots show the same under a frequentist point prior on $p_1=0.4$ under $H_1$. Flat analysis priors are assumed in both cases.}
    \label{fig:ex3}
\end{figure}

In this subsection, we provide a quick demonstration of what gains can be expected when shifting from a non-sequential Bayes factor design to our novel sequential design. Thus, we compare a Bayesian power analysis without an interim analysis with the proposed optimal sequential Bayes factor design with a single interim analysis at $n_1$. The general approach to carry out a Bayes factor design analysis without any interim analyses has been detailed by \cite{KelterPawel2025}. \Cref{fig:ex3} shows the results for the second example above, where in the phase II trial the hypotheses $H_0:p\leq 0.2$ and $H_1:p>0.2$ were under consideration.

\Cref{fig:ex3_1} provides the sample sizes $n=110$ for $k=1/10$ and $n=61$ for $k=1/3$ under a flat Beta design prior on $[0.2,1]$ under $H_1$. As there is no interim analysis, this is comparable to the full sample size $n_2$ in the optimal sequential Bayes factor designs in \Cref{tab:ex2}. The thresholds for power and type-I-error rates are identically set to $90\%$ and $10\%$. \Cref{tab:ex2} shows that the sequential design reduces $n=110$ to $n_2=100$ with the option to stop early at $n_1=42$. The expected sample size is $E[N|H_0]=69.21$, about $40$ patients less than the fixed $n=110$ in the non-sequential Bayes factor design analysis, which always requires $n=110$ patients. 

The right panel of \Cref{fig:ex3_1} shows that when using the evidence threshold $k=1/3$ for moderate evidence, the required maximum sample size is $n=61$. \Cref{tab:ex2} shows that this sample size reduces to $n=54$ when using the two-stage Bayes factor design, and the expected sample size becomes $E[N|H_0]=39.46$ under the null hypothesis. Thus, a reduction of the required number of patients of about a third is possible when shifting from the non-sequential to the proposed two-stage design.

\Cref{fig:ex3_2} provides analogue results under a point prior at $p_1=0.4$, so power is calculated in a fully frequentist sense. The resulting sample sizes $n=53$ for $k=1/10$ and $n=36$ for $k=1/3$ can be compared to the ones in \Cref{tab:ex2} again. For $k=1/3$, the results in \Cref{fig:ex3_2} can be compared to the ones for $k=1/3$ and $k_f=3$ in \Cref{tab:ex2}, which shows that the expected sample size in the sequential design is $E[N|H_0]=26.02$, which is the one of Simon's minimax design.\footnote{This is reasonable, because the approach in \cite{KelterPawel2025} searches for the smallest sample size which fulfills the requirements in \Cref{eq:bayesianDesign}, except without any interim analysis.} Thus, shifting to the sequential design reduces the required sample sizes significantly about a third. For $k=1/10$, the optimal sequential Bayes factor design with $k=1/10$ and $k_f=3$ yields a design which has an expected sample size of $E[N|H_0]=33.42$, about $20$ patients less than in the non-sequential design, which needs $n=54$ patients according to the left panel in \Cref{fig:ex3_2}.

Expected sample size under $H_0$ is reduced similarly in other scenarios, when shifting to the sequential Bayes factor design based on our idea of trinomial tree branching.

\section{Discussion}\label{sec:discussion}
Sequential trial design is an important statistical approach to increase the efficiency of clinical trials. Bayesian sequential trial design relies primarily on conducting a Monte Carlo simulation under the hypotheses $H_0$ and $H_1$ of interest and the resulting design characteristics are then investigated by means of Monte Carlo estimates. This approach has several drawbacks, namely that calibration of a Bayesian design requires repeating a possibly complex Monte Carlo simulation. This holds both for the process of finding a calibrated design itself, and also for replicating a calibrated design to check the validity of the proposed design (which is relevant for regulatory agencies). In addition to the time needed for these two points, Monte Carlo standard errors are required to judge the reliability of the simulation.

All of this is due to the fact that closed-form or numerical approaches to calibrate a Bayesian design are essentially missing from the literature. In this paper, we therefore proposed the Bayesian optimal two-stage design for clinical phase II trials based on Bayes factors. Therefore, we developed the approach of trinomial tree branching, a method to visualize and illustrate the challenges which occur during power and type-I-error-rate calculations when introducing a single interim analysis into a static design, thereby making it sequential. Next, we developed theoretical results which allowed to correct the resulting design characteristics such as power and type-I-error rate for introducing an interim analysis. 

We build upon this idea to invent a simple calibration algorithm and an optimal calibration algorithm, which yields the optimal Bayesian design that minimizes the expected sample size under the null hypothesis $H_0$. This is inspired from the well-established Simon's two-stage optimal design for phase II trials with a binary endpoint. Illustrative examples showed that our design recovers Simon's two-stage optimal design as a special case, improves upon non-sequential Bayesian design based on Bayes factors, and can be calibrated very fast with computation times usually below a minute on a regular desktop computer. This is, in turn, due to the almost instant computation of the power and type-I-error rate of our design as developed in \Cref{sec:trinomial}. The last aspect, in particular, makes it a highly attractive choice for application in practice. Comparisons of multiple calibrated designs like in \Cref{tab:ex_oscillations} and \Cref{tab:ex2} -- with possibly different priors, power concepts (Bayesian or frequentist), evidence thresholds or futility thresholds -- are easily carried out within seconds. Furthermore, the design allows to guarantee a minimum probability on compelling evidence in favour of the null hypothesis, which is that easily possible with other designs.

We close this discussion with a few points which are relevant and some venues for future work:
\begin{itemize}
    \item[$\blacktriangleright$]{The core idea of trinomial tree branching that underpins the calibration algorithm is neither dependent on the binary endpoint, nor on the use of Bayes factors, nor the choice of priors. As a consequence, the design can principally be generalized to other settings with a different endpoint, a different testing approach (e.g. posterior probability, posterior odds), and different priors. This provides a profound branch of future work, namely to develop Bayesian optimal two-stage designs for other settings. For example, a natural extension would be to shift from a single-arm phase II trial to a two-arm phase II trial with binary endpoints (in treatment and control group). Three key preliminaries are then given as follows:
    \begin{itemize}
        \item{First, when shifting to different endpoints, the Bayes factor design analysis provided by \cite{KelterPawel2025} must be modified accordingly. This enables to compute the power and type-I-error rate for a fixed sample size, compare \Cref{sec:calibration}. \cite{PawelHeld2025} provide results how to do this for a variety of endpoints, so this should not present too much of a hurdle for extending the current design to other settings.}
        \item{Second, \Cref{theorem:1} makes use of the marginal probability mass function $$f(y_s^1,y_s^2|a_d,b_d,l,u)$$ given the design prior on $H_1$. The second preliminary to extend the proposed design therefore is to be able to compute this marginal probability mass (or density, for continuous endpoint(s)) function at least numerically.}
        \item{Third, the test criteria of choice must be computable analytically or at least numerically. So, if for example the case of a two-arm phase II trial with binary endpoints is considered, a different Bayes factor is required. However, Bayes factors are available in the literature for most practically relevant settings by now \citep{VanRavenzwaaij2019,Du2019,Morey2011,Rouder2009,Ly2016a,Rouder2012}. For other test criteria like posterior probabilities or posterior odds, these can also be computed at least numerically in most cases.}
    \end{itemize}}
    \item[$\blacktriangleright$]{We did not explicitly incorporate the idea to assert a minimum probability for convincing evidence in favour of $H_1$ when running the interim analysis at $n_1$. Also, we did conceptually not allow the trial to be stopped for efficacy at $n_1$. However, a possible generalization of our design might include such an option.}
    \item[$\blacktriangleright$]{We provided an optimal calibration algorithm which makes use of the early ideas of \cite{Simon1989} and which minimizes the expected sample size $E[N|H_0]$ under the null hypothesis $H_0$, compare the discussion in \Cref{sec:calibration}. Still, future work could consider different calibration routines such as minimax calibration or even minimin calibration (when carrying out the interim analysis as early as possible is desired). When taking into account \Cref{eq:futilityInterim}, even calibration algorithms would be possible which maximize the probability of compelling evidence for the null hypothesis under other constraints like an upper boundary $n\in \mathbb{N}$ for $E[N|H_0]$:
    \begin{equation}
        \begin{aligned}
            \max_{n_1,n_2} \quad & P(\mathrm{BF}_{01}^{n_1}(y^1)>1/k|H_0)\\
            \textrm{s.t.} \quad & E[N|H_0]\leq n\\
              &n_{min}\leq n_1 < n_2 \leq n_{max}    \\
        \end{aligned}
    \end{equation}}
\end{itemize}
The above shows that there exist plenty of possibilities to extend and generalize the proposed Bayesian optimal two-stage design for clinical phase II trials based on Bayes factors. Given the ease of application and the versatility of our proposed sequential design, we hope that it will be a useful addition to the toolkit of medical statisticians faced with the task of designing a single-arm phase II trial with binary endpoint.

\appendix
\appendixpage

\section*{Proofs}\label{sec:proofs}
\begin{proof}[Proof of \Cref{theorem:1}]
    \begin{align}\label{eq:derivPartialPower}
    P(\mathrm{BF}_{01}^{n_2}<k,\mathrm{BF}_{01}^{n_1}>1/k \mid H_i)&=P(y_s^2>y_{critEff}^{n_2}-y_s^1,y_s^1 \leq y_{critFut}^{n_1}|H_i)\nonumber\\
    &=\int_{\{y_s^1 \leq y_{critFut}^{n_1}, y_s^2>y_{critEff}^{n_2}-y_s^1\}}1dP_{H_i}\nonumber\\
    &=\int_{\{y_s^1 \leq y_{critFut}^{n_1}, y_s^2>y_{critEff}^{n_2}-y_s^1\}}f(y_s^1,y_s^2|H_i)dy\nonumber\\
    &=\int_{\{y_s^1 \leq y_{critFut}^{n_1}, y_s^2>y_{critEff}^{n_2}-y_s^1\}}f(y_s^1, y_s^2|a_d,b_d,l,u)dy\nonumber\\
    &=\sum_{\{y_s^1 \leq y_{critFut}^{n_1}, y_s^2>y_{critEff}^{n_2}-y_s^1\}}f(y_s^1, y_s^2|a_d,b_d,l,u)dy\nonumber\\
    &=\sum_{i=0}^{\lfloor y_{critFut}^{n_1}\rfloor}\sum_{j=\lceil y_{critEff}^{n_2}\rceil -y_i}^{n_2-n_1} f(y_s^1=y_i, y_s^2=y_j|a_d,b_d,l,u)dy\nonumber\\
\end{align}
In the above, we first make use of $\mathrm{BF}_{01}^{n_2}(y)<k$ if and only if $y_s^2>y_{critEff}^{n_2}$, where $y_s^1$ and $y_s^2$ denote the number of successes observed in $y^1$ and $y^2$ at sample size $n_1$ and $n_2$. Thus, the Bayes factor $\mathrm{BF}_{01}^{n_2}(y)$ based on the full data of sample size $n_2$ shows evidence $k$ against $H_0$ if and only if we observe at least $y_{critEff}^{n_2}$ successes in the full trial data $y$. If we have already observed $y_s^1$ successes in the first batch of interim data $y^1:=(y_1,...,n_1)$, then we need only $y_{critEff}^{n_2}-y_s^1$ successes to yield a Bayes factor $\mathrm{BF}_{01}^{n_2}(y)<k$ in the final analysis based on the full data $y:=(y^1,y^2)=(y_1,...,y_{n_1},y_{n_1+1},...,y_{n_2})$. An analogue holds for $\mathrm{BF}_{01}^{n_1}(y)>1/k$.

Next, we make use of $$P(y_s^2>y_{critEff}^{n_2}-y_s^1 , y_s^1 \leq y_{critFut}^{n_1} | H_i)=\int_{\{y_s^2>y_{critEff}^{n_2}-y_s^1, y_s^1 \leq y_{critFut}^{n_1}\}}1dP_{H_i}$$
where $P_{H_i}$ denotes the conditional probability measure $P(\cdot |H_i)$ on the hypothesis $H_i$. Then, we apply rewriting the Radon-Nikodym-density $dP_{H_i}/dy=f(y_s^1, y_s^2|H_i)=f(y_s^1, y_s^2|a_d,b_d,l,u)$ as we use truncated beta design priors $\theta \sim \mathrm{Beta}(a,b)_{[l,u]}$, for details see below. Also, we make use of the fact that the integral reduces to a discrete sum, which can be expressed as a summation over two variables $i$ and $j$. In these sums, we need to use floor and ceiling functions, as we need to sum over all $y_s^1 \leq y_{critFut}^{n_1}$ (so that we need to use a floor function on $y_{critFut}^{n_1}$), and over all $y_s^2>y_{critEff}^{n_2}-y_s^1$ (so we need to use a ceiling function on $y_{critEff}^{n_2}$). Importantly, in the inner sum we can replace $\lceil y_{critEff}^{n_2}\rceil -y_s^1$ with $\lceil y_{critEff}^{n_2}\rceil -y_i$ then, as $y_i$ counts the number of successes $y_i=0,...,\lfloor y_{critFut}^{n_1} \rfloor$ in the first batch of data in the outer sum.

To make use of the above representation of the marginal probability mass function, it suffices to note that
if $Y_i \mid \theta \sim \mathrm{Bin}(n_i, \theta)$ for $i\in\{1,2\}$ and $\theta \sim \mathrm{Beta}(a,b)_{[l,u]}$, then the marginal probability mass function $f(y_s^1,y_s^2|a_d,b_d,l,u)$ is given as
\begin{align*}
    f(y_s^1,&y_s^2|a_d,b_d,l,u) 
    = \int_l^u f(y_s^1 \mid \theta) f(y_s^2 \mid \theta) f(\theta|a_d,b_d) \mathrm{d}\theta \\
    &= \int_l^u \binom{n_1}{y_s^1} \theta^{y_s^1} (1 - \theta)^{n_1-y_s^1} \binom{n_2}{y_s^2}  \theta^{y_s^2} (1 - \theta)^{n_2-y_s^2} \frac{\theta^{a_d-1} (1 - \theta)^{b_d-1}}{\mathrm{B}(a_d,b_d) \{I_u(a_d, b_d) - I_l(a_d, b_d)\}} \mathrm{d}\theta \\
    &= \binom{n_1}{y_s^1} \binom{n_2}{y_s^2} \frac{\mathrm{B}(a_d + y_s^1 + y_s^2,b_d + n_1 + n_2 - y_s^1 - y_s^2)}{\mathrm{B}(a_d,b_d)\{I_u(a_d, b_d) - I_l(a_d, b_d)\}} \\
    &\times \int_l^u \frac{\theta^{a_d + y_s^1 + y_s^2 - 1} (1 - \theta)^{b_d + n_1 + n_2 - y_s^1 - y_s^2 - 1}}{\mathrm{B}(a_d + y_s^1 + y_s^2,b_d + n_1 + n_2 - y_s^1 - y_s^2)} \mathrm{d}\theta \\
    &= \binom{n_1}{y_s^1} \binom{n_2}{y_s^2} \frac{\mathrm{B}(a_d + y_s^1 + y_s^2,b_d + n_1 + n_2 - y_s^1 - y_s^2)}{\mathrm{B}(a_d,b_d)\{I_u(a_d, b_d) - I_l(a_d, b_d)\}} \\
    &\phantom{=} \times \{ I_u(a_d + y_s^1 + y_s^2, b_d + n_1 + n_2 - y_s^1 - y_s^2) - I_l(a_d + y_s^1 + y_s^2, b_d + n_1 + n_2 - y_s^1 - y_s^2)\}
\end{align*}
so calculating the marginal probability mass function $f(y_s^1,y_s^2|a_d,b_d,l,u)$ is possible with standard numerical integration.
\end{proof}

\bibliography{library.bib} 

\end{document}